\documentclass{aa}  
\usepackage{graphicx,url,twoopt,natbib}
\usepackage[varg]{txfonts}           
\usepackage{hyperref}                

\hypersetup{
  colorlinks=true,   
  urlcolor=blue,     
  linkcolor=blue     
}

\bibpunct{(}{)}{;}{a}{}{,}    
\makeatletter
\newcommand{\bibnote}[2]{\global\@namedef{#1note}{#2}}
\newcommand{\biblink}[2]{\global\@namedef{#1link}{#2}}
\makeatother

\makeatletter
  \protected\def\stonyslink{%
     \def\hyper@linkstart##1##2{}\let\hyper@linkend\@empty}
  \newcommandtwoopt{\citeads}[3][][]{%
   \href{http://ui.adsabs.harvard.edu/abs/#3/abstract}%
        {\stonyslink \citealp[#1][#2]{#3}}
   \biblink{#3}{\href{http://ui.adsabs.harvard.edu/abs/#3/abstract}{ADS}}}
 \newcommandtwoopt{\citepads}[3][][]{%
   \href{http://ui.adsabs.harvard.edu/abs/#3/abstract}%
        {\stonyslink \citep[#1][#2]{#3}}
   \biblink{#3}{\href{http://ui.adsabs.harvard.edu/abs/#3/abstract}{ADS}}}
 \newcommandtwoopt{\citetads}[3][][]{%
   \href{http://ui.adsabs.harvard.edu/abs/#3/abstract}%
        {\stonyslink \citet[#1][#2]{#3}}
  \biblink{#3}{\href{http://ui.adsabs.harvard.edu/abs/#3/abstract}{ADS}}}
 \newcommandtwoopt{\citeyearads}[3][][]{%
   \href{http://ui.adsabs.harvard.edu/abs/#3/abstract}%
        {\stonyslink \citeyear[#1][#2]{#3}}
   \biblink{#3}{\href{http://ui.adsabs.harvard.edu/abs/#3/abstract}{ADS}}}
\makeatother

\begin{document}  
\nolinenumbers
\title{Morphological evidence for nanoflares heating warm loops in the solar corona
  \thanks{Movies associated to Figs. 3, 4, and 5are available. }}
\author{Yi Bi\inst{1,2}
  \and Jia-Yan Yang\inst{1}
     \and Ying Qin\inst{1,2}
     \and Zheng-Ping Qiang\inst{3}
       \and Jun-Chao Hong\inst{1}
  \and Bo Yang\inst{1}     
   \and Zhe Xu\inst{1} 
    \and Hui Liu\inst{1,2}
      \and Kai-Fan Ji\inst{1}          }
\institute{Yunnan Observatories, Chinese Academy of Sciences, 396 Yangfangwang, Guandu District, Kunming, 650216, People's Republic of China \\\email{biyi@ynao.ac.cn,jkf@ynao.ac.cn}
  \and Yunnan Key Laboratory of Solar Physics and Space Science, 396 Yangfangwang, Guandu District, Kunming 650216, People’s Republic of China
  \and College of Big Data and Intelligent Engineering, Southwest Forestry University, Kunming 650224, China}

\abstract {Nanoflares are impulsive energy releases by magnetic reconnection in the braided coronal magnetic field, which is a potential mechanism for heating the corona. However, there are still sporadic observations of the interchange of braiding structure segments and footpoints inside coronal loops, which is predicted to be the morphological evolution of the reconnecting magnetic bundles in the nanoflare picture. }
 {This work aims to detect  the evolutions of the pairs of braiding strands within the apparent single coronal loops observed in Atmospheric Imaging Assembly (AIA) images. }  {The loop strands are detected  on  two kinds of  upsampled AIA 193 \AA\ images, which are  obtained by upscaling the Point Spread Function matched AIA images via Bicubic interpolation and are  generated using a super-resolution convolutional neural  network, respectively. The architecture of the network is designed to map the AIA images to unprecedentedly high spatial resolution coronal images taken by  High-resolution Coronal Imager (Hi-C) during its brief flight. } 
{At times, pairs of separate strands that appear braided together later evolved into pairs of almost parallel strands with completely exchanged parts. These evolutions offer morphological evidence that magnetic reconnections between the braiding strands have taken place, which is further supported by the appearance of transient hot emissions containing significant high-temperature components (T > 5MK) at the footpoints of the braiding structures. } 
{The brief appearances of the two rearranging strands support that magnetic reconnections have occurred within what appears to be a single AIA loop.}

\maketitle

\section{Introduction}     \label{sec:introduction}

\nolinenumbers
One of the most challenging problems in solar physics is how the solar corona is heated up to  a temperature of millions of degrees, far above that of the photosphere, although it is widely accepted that 
 the magnetic field plays a major role in the energetics of the bright corona.
 
 The bright  coronal loops are the building blocks of the solar corona. Therefore the heating mechanisms for the coronal loops are important to understand how the corona is heated.
Based on the temperature regime, the loops observed in EUV are classified as warm loops and hot loops (\citeads{Reale14}), which confine plasma at temperatures around 1-1.5 MK and around or above 2 MK, respectively.
The model developed by \citetads{Van17} indicated that the   Alfv{\'e}n wave turbulence launched from the photosphere  can produce enough heat to maintain a peak temperature of about 2.5 MK of the coronal loops.
 Also, a large number of transverse waves are deduced from the observed Alfv{\'e}nic motion of the coronal features (\citeads{McIntosh}),  spicules (\citeads{DePontieu}), and network jets (\citeads{Tian14,Shen}), as well as the falling solar prominence knots (\citeads{Bi}).

Energy releases from small-scale magnetic reconnections are another promising mechanism to heat the corona  (\citeads{Klimchuk}).
It has been accepted that the small-scale events of magnetic reconnections (\citeads{Testa13,Gupta,Priest,Asgari19,Chitta20})  are responsible for the  heating of the hot loops  or hot plasma in the corona (\citeads{Klimchuk,Schmelz,Ishikawa,Yang18,Zhang23}).
It seems more controversial  how the warm loops are heated.
Some warm loops might be globally cooling from the hot loops (\citeads{Winebarger05,Viall,Li}), but many long-lived warm loops  would be much less visible in the hot EUV channels.
Although Alfv{\'e}n waves originating in the photosphere may provide sufficient energy for heating the warm loops (\citeads{Van17}), 
observations supporting the  magnetic reconnection-type heating in the warm loops were also presented, such as the transient brightnesses found around the footpoints of the warm loops (\citeads{Regnier14,Subramanian}), short-lived warm loops  that impulsively appeared in and faded out during a few minutes and never achieve million-degree temperatures (\citeads{Winebarger}), and the reconnection outflow-like plasma (termed nanojet) within the warm loops composed of misaligned strands  (\citeads{Antolin}).

    A nanoflare refers to an impulsive energy release in the coronal braided magnetic field (\citeads{Parker88}), which are considered the most promising mechanism for the generation of hot plasma in active regions by small-scale reconnection.  According to the nanoflare scenario, when the strands reconnect, they exchange segments and footpoints (\citeads{Berger,Klimchuk15}).    However, the morphological evolutions of sub-arcsecond strands linked to nanoflares are still challenging to identify, probably due to the existing performance limits of coronal observations  and the scarcity of coronal observations with spatial resolution less than 1{\arcsec}, such as  High-resolution Coronal Imager (Hi-C; \citeads{Kobayashi}) and Extreme Ultraviolet Imager (EUI).
    
    It has been reported that  two braiding structures (\citeads{Cirtain}) were detected in the  images taken by the Hi-C,  which observed a bandpass dominated by the Fe xii 193 \AA\ line with a pixel size of $\sim$ 0{\arcsec}.1 ($\sim$ 75 Km on the Sun) in the period of a few minutes.
    Using NLFFF extrapolation, \citetads{Thalmann} have found that the braided structure observed by Hi-C was a low-lying twisted flux
rope above a penumbral filament region.
However,  limited to a  brief period of observation,  Hi-C is unable to confirm whether these tangled loops are associated with energy release in the solar corona.
 Most recently,  using  new observations with high spatial resolution (a pixel size of 125 -135 km on the Sun) from  EUI on board Solar Orbiter, 
\citetads{Chitta22} reported the  untangling of small-scale coronal braids, giving rise to coronal loops that run more parallel with each other. 
By contrast, the uninterrupted observations taken by Atmospheric Imaging Assembly (AIA; \citeads{Lemen}) onboard the Solar Dynamics Observatory (SDO) 
have a larger pixel size of   $\sim$ 0{\arcsec}.6  and  then hardly recognize the  braiding substructures as resolved in the Hi-C and EUI images.

  Recently, various machine learning (ML) models have been applied to create artificial solar images
     to further extend the performance of the current solar observations (\citeads{Kim,Szenicer,Bai,Hong,Dos_santos,Pineci,Yu21}).
   In particular, \citetads{Dazbaso}  adopted a deep neural network approach  to   deconvolve and  supperresolve Helioseismic and Magnetic Imager (HMI; \citeads{Schou}) images and found that the synthetic HMI images  contained information not present in the original data, which supported that a certain deep convolutional neural network could be applied to  solar image enhancement. Super-Resolution (SR) is a classic problem in computer vision, which is to recover a high-resolution image from a low-resolution image. Because there are multiple solutions for any given low-resolution pixel, SR is generally an ill-posed inverse problem. Its solution pipeline is suggested to be  equivalent to a deep convolutional neural network (\citeads{Dong}), 
  and then a proper architecture of a convolutional neural network could be applied to directly learn an end-to-end mapping between low- and high-resolution images.

The rest of the paper is structured as follows.
Section \ref{s2} introduces the details of the algorithm for upscaling the AIA 193 \AA\ images.  Here, the images are upscaled by 
the ML-based mapping from AIA to Hi-C 193 \AA\  images,
  as well as by the method of deconvolution and upscaling interpolation. 
Section \ref{s3} presents three events not covered by  Hi-C observations, in which the reconnection-like  rearrangements of the braiding strands are observed on the upscaled   193 \AA\ images and the transient hot emissions are detected on the AIA 94 \AA\ images.  Discussions and a brief summary are presented in Sects.  \ref{s4} and \ref{s5}.

\begin{figure}[hbtp]
  \centering
  \includegraphics[width=\columnwidth]{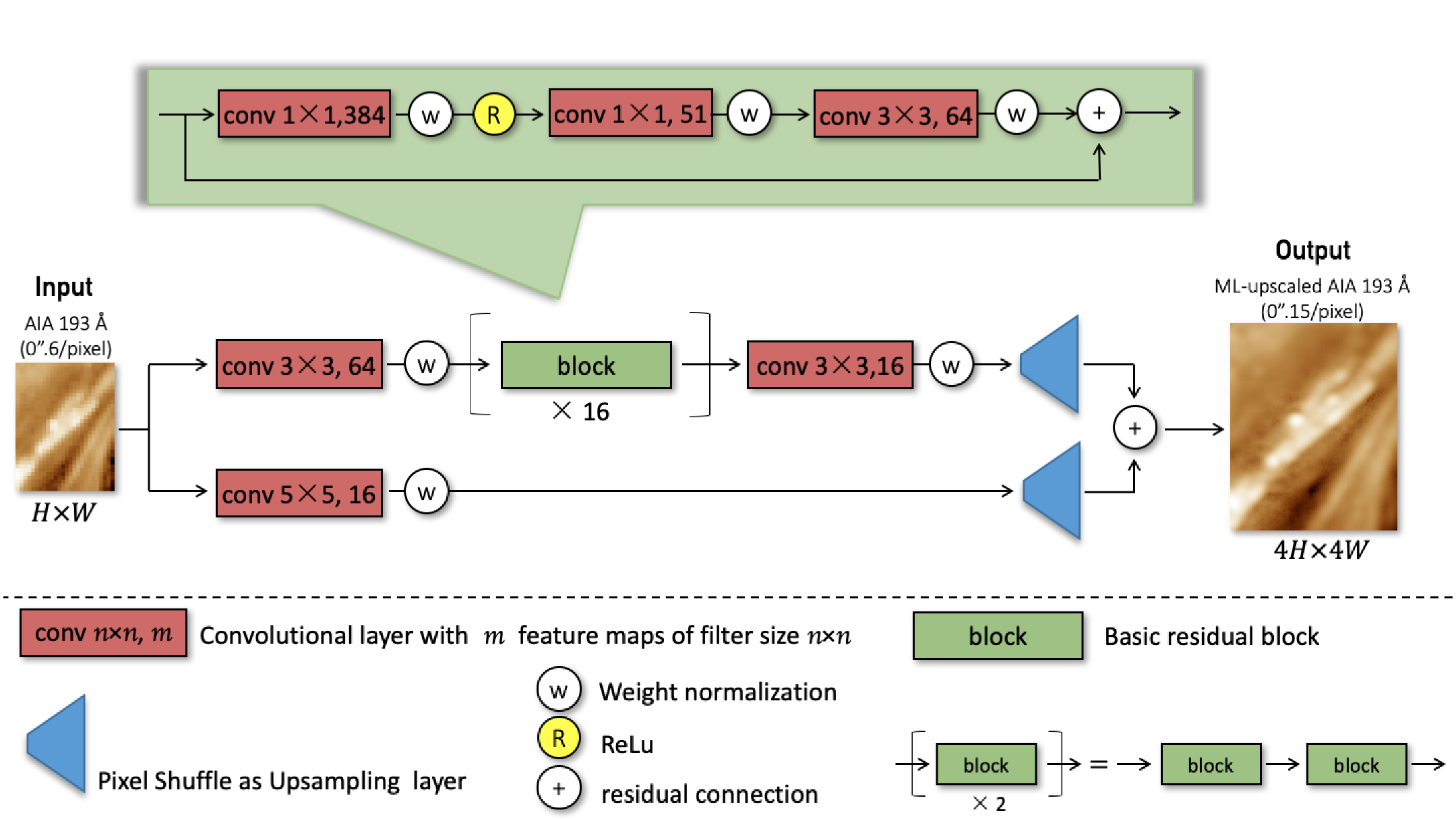}  
  \caption[]{\label{f1} 
 Architecture of the WDSR convolutional neural network  for upsampling factor $\times 4$.
All of feature maps in each convolutional layer have the size $H \times W$ as same as that of input image/patch. 
For upscaling factor $\times S$ ($\times 4$), the number of channels used as the input for the Pixel-Shuffle layer is exactly $S^2$ (16).
}
\end{figure}

\begin{figure*}[hbtp]
  \resizebox{\hsize}{!}{\includegraphics{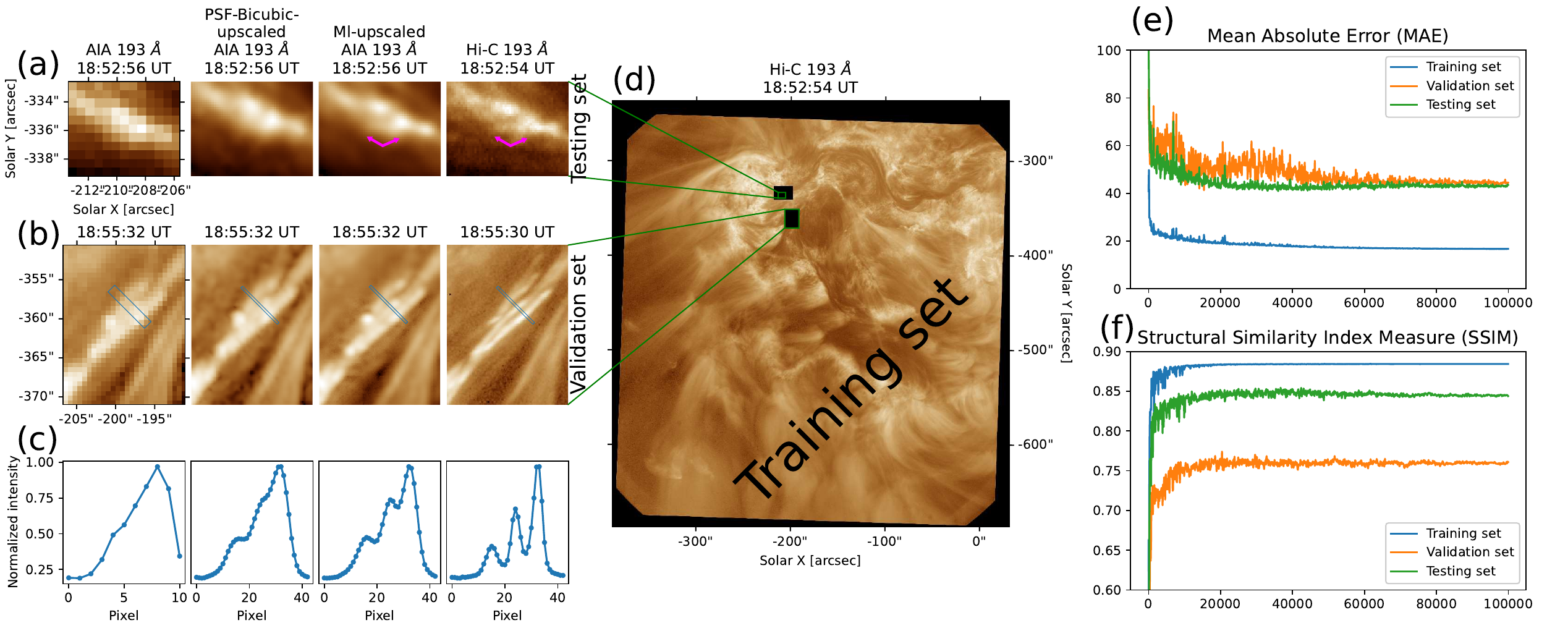}}
  \caption[]{\label{f2} 
  The training, testing, and validation sets for the machine learning. 
{\em Panels a-b:\/} Comparison of close-ups of  the unprocessed AIA, PSF-Bicubic upscaled AIA, Ml upscaled AIA, and Hi-C images.
The images in ({\em a}) and ({\em b}) are taken  from testing set and  validation set, respectively. 
{\em Panels c:\/}   The slice plots exhibiting intensities  along the line  indicated  in {\em b}.
{\em Panel d:\/}  The Hi-C image with its full field-of-view, in which the lower and upper blanked regions are taken as the validation set and testing set, respectively, and all of the left region is used as the training set.
{\em Panels e-f:\/}  
Time history of the value of MAE ({\em e}) and SSIM ({\em f}) between the model outputs and the Hi-C images in each dataset.}
\end{figure*}
\section{Method}    \label{s2}

 We use two methods to upscale the AIA  193  \AA\ images to the  pixel  size of 0{\arcsec}.15.
Firstly, the AIA images are deconvolved with the  Point Spread Function (PSF\footnote{The PSF for AIA is available in SolarSoft (\url{http://www.lmsal.com/solarsoft/}).}; \citeads{Boerner})  via Richardson-Lucy algorithm, and then the resulting PSF-matched images are upsampled using Bicubic interpolation. We term them  as PFS-Bicubic upscaled images.
Secondly, the  ML-upscaled AIA  193 \AA\ images are generated using a super-resolution network mapping the AIA 193 \AA\ images to Hi-C images.

 To generate the ML-upscaled AIA 193 \AA\ images, we applied  the wide activation  super-resolution   networks (WDSR;  \citeads{Yu,Fan2018}), a kind of ML network for single image super-resolution  (SR), to  upsample the AIA 193 \AA\ images by a factor of 4.
As shown in Fig. \ref{f1}, the network mainly consists of 16 residual blocks (\citeads{He}).
 Each basic block is made up of three convolutional layers and starts with the feature maps being extended to 384  with $1 \times 1$ kernel. 
The channel number expansion before the ReLU pooling layer, termed  wide  activation  (\citeads{Yu}), allows more information to pass from shallow layers to deeper ones.
 A global residual connection is applied to relieve  the redundant features generated potentially by the deep network architectures (\citeads{Ledig}).
 A pixel-Shuffling  layer (\citeads{Shi}) is utilized at the end of the network to upsample  the final  feature maps into the SR output.
 Weight Normalization layers (\citeads{Salimans}) are used to ease the training difficulty of deep networks.  The total number of trainable parameters for the network modeling amounts to $1.2 \times 10^{6}$.

Five pairs of AIA and Hi-C images are used to train the ML-upscaled model.
The observed time difference between each pair of images is less than 3 seconds.
 Using these images we built the training set and validation set as follows.
 Firstly, the  Hi-C images with a pixel size of 0{\arcsec}.1  are downsampled via bilinear interpolation to  a pixel size of 0{\arcsec}.15,  a quarter of the pixel size of the AIA images.
Secondly, we aligned the Hi-C to AIA images via cross-correlation.
 Finally,   two small patches (indicated by the lower and upper blanked regions in Fig. \ref{f2}d) in these images are extracted as  the validation  and testing  sets. 
The training set is made up of all frames of the data  with the validation and test sets being spared out. That is,  to prevent the model to train the known state of that regions shortly before/after that frame, it is necessary to ensure that these regions in the validation and testing set are not taken for training from any frame.

The frame patches for training have  a size of 48 $\times$ 48 pixels, which are randomly  extracted both spatially and temporally from the training set and are randomly rotated in multiples of 90 degrees.
The number of patches  used in one single training epoch amount to 100 and the number of epochs used for training is  $10^5$. 
 We set an initial learning rate of $1.2 \times 10^{-5}$, decreased by 90 \% every 2000 epochs and use the loss function of   the mean absolute error (MAE)  function for optimizing the network. The code and trained model are based on Pytorch and available at GitHub\footnote{\url{https://github.com/YiBi-YNAO/ML-upscaled-AIA-193-.git}}.

  For  training, validation,  and testing sets,  the quantities of MAE between the ML-upscaled and Hi-C images (ground truth) are seen to be  flattened out finally  after about $5 \times 10^{4}$ training epochs are performed (Fig. \ref{f2}e).
  This indicates that the model has not been overfitted.
To further evaluate how well our model enhances the AIA 193 \AA\ images, we use Structural Similarity Index Measure (SSIM; \citeads{Wang04,Aydin})   for measuring similarity between  the ML-upscaled and Hi-C images. A greater value of SSIM reflects a smaller difference between them.
    As expected, the value of SSIM in  each set  increased initially during training and was then close to an asymptotic state (Fig. \ref{f2}f).
Figs.  \ref{f2}a and \ref{f2}b compares the unprocessed AIA, PSF-Bicubic upscaled AIA, Ml upscaled AIA, and Hi-C images in the testing and validation sets, respectively.
On the Hi-C image, an apparent single AIA loop in the validation set appears to be three separate strands, which were identified by \citetads{Cirtain} as a set of braiding loops.
The split strands were clearly visible in the ML-upscaled AIA images, as well as slightly in the PSF-Bicubic-upscaled AIA images (Fig. \ref{f2}b). 
Fig. \ref{f2}c shows the slice plots across the loop in the various images, with three peaks evident in the slice plots taken from the ML-upscaled and Hi-C image. 
 In the testing set, again, a bifurcation of an AIA loop could be seen on  both ML-upscaled AIA and Hi-C images (as indicated by the arrows in Fig. \ref{f2}a).
Thus, while both the PSF-Bicubic and ML upscaling methods are capable of improving the AIA 193 \AA\ images in a reasonable manner, the ML upscaling method appears to perform better in both the validation and testing sets.

\begin{figure*}[hbtp]
  \centering
    \includegraphics[width=1.4\columnwidth]{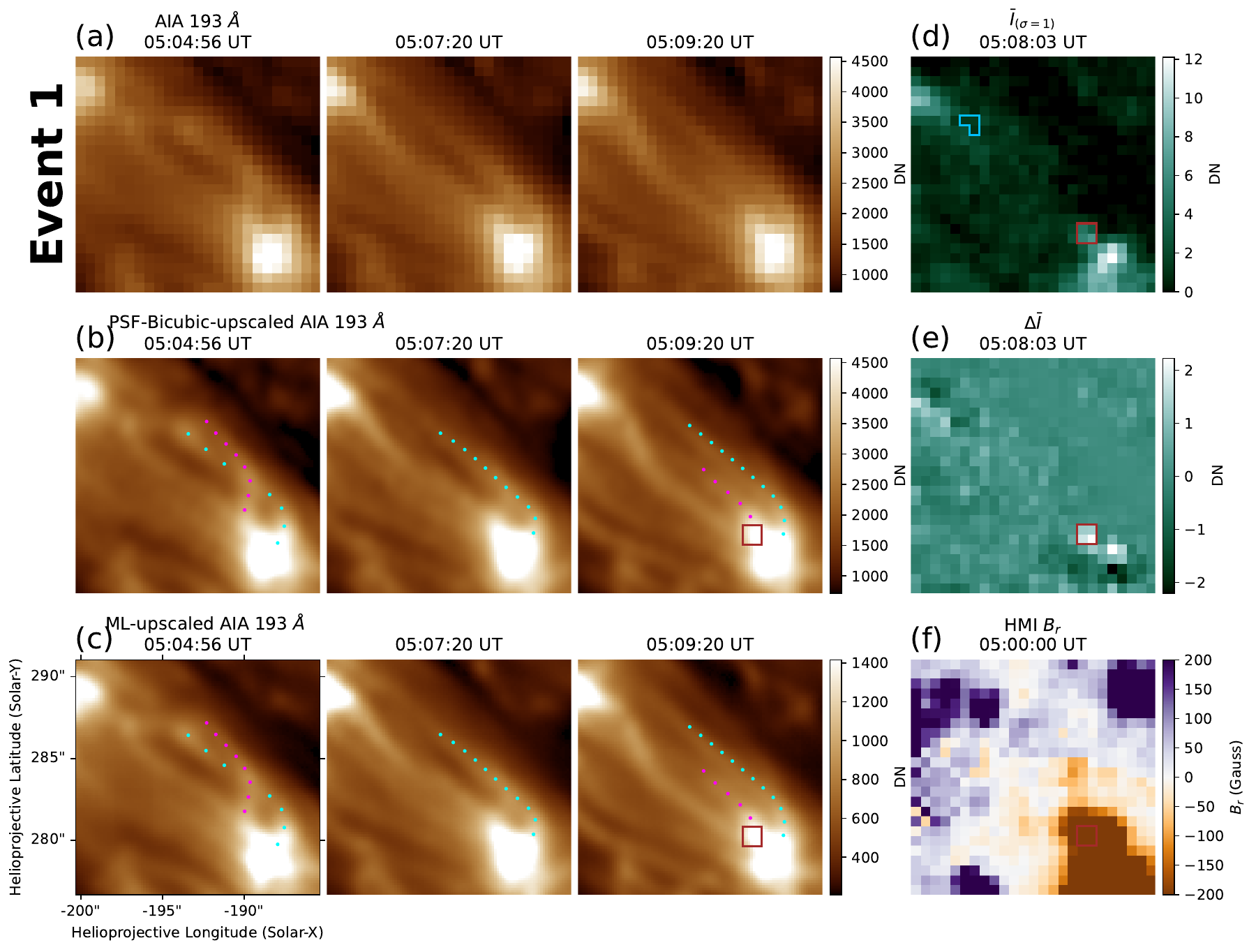}
  \caption[]{\label{f3} 
The evolution of the braiding strands on  5 January 2012 (Event 1).
{\em Panel a:\/}  SDO/AIA 193 \AA\ images.  
 {\em Panels b-c:\/}  the Bicubic upscaled versions of the   PSF-matched AIA 193 \AA\ images 
 and ML-upscaled versions of the AIA 193 \AA\ images, on which  the various colored dot-lines indicate the various recognizable  strands.
  {\em Panels d-e:\/}  The images of $\bar{I}_{(\sigma=1)}$ and  $\Delta \bar{I}=\bar{I}_{(\sigma=1)}-\bar{I}_{(\sigma=4)}$, where the $\bar{I}_{(\sigma=1)}$ and $\bar{I}_{(\sigma=4)}$ is the Fe XVIII intensity of  $I(94 \AA\ )-I(211 \AA\ )/120-I(171 \AA\ )/450$ smoothed with a Gaussian kernel of 1 and 4 minutes, respectively.
  These  images  are taken at the peak time of the hot emission  indicated by the brown box in {\em d}, which is also plotted on the other panels.
The boxes in {\em d} outline   the locations of all  hot emissions that are  identified from 05:00:08 UT to 05:10:08 UT. 
{\em Panel f:\/}  The vertical component  ($B_{r}$) of the photospheric  vector data from SDO/HMI.
   All images have the same Field-Of-View (FOV). 
   The evolution of the braiding strands is shown in a movie (anim3.mpeg) available online.
}
\end{figure*}

   \begin{figure*}[hbp]
  \centering
      \includegraphics[width=1.4\columnwidth]{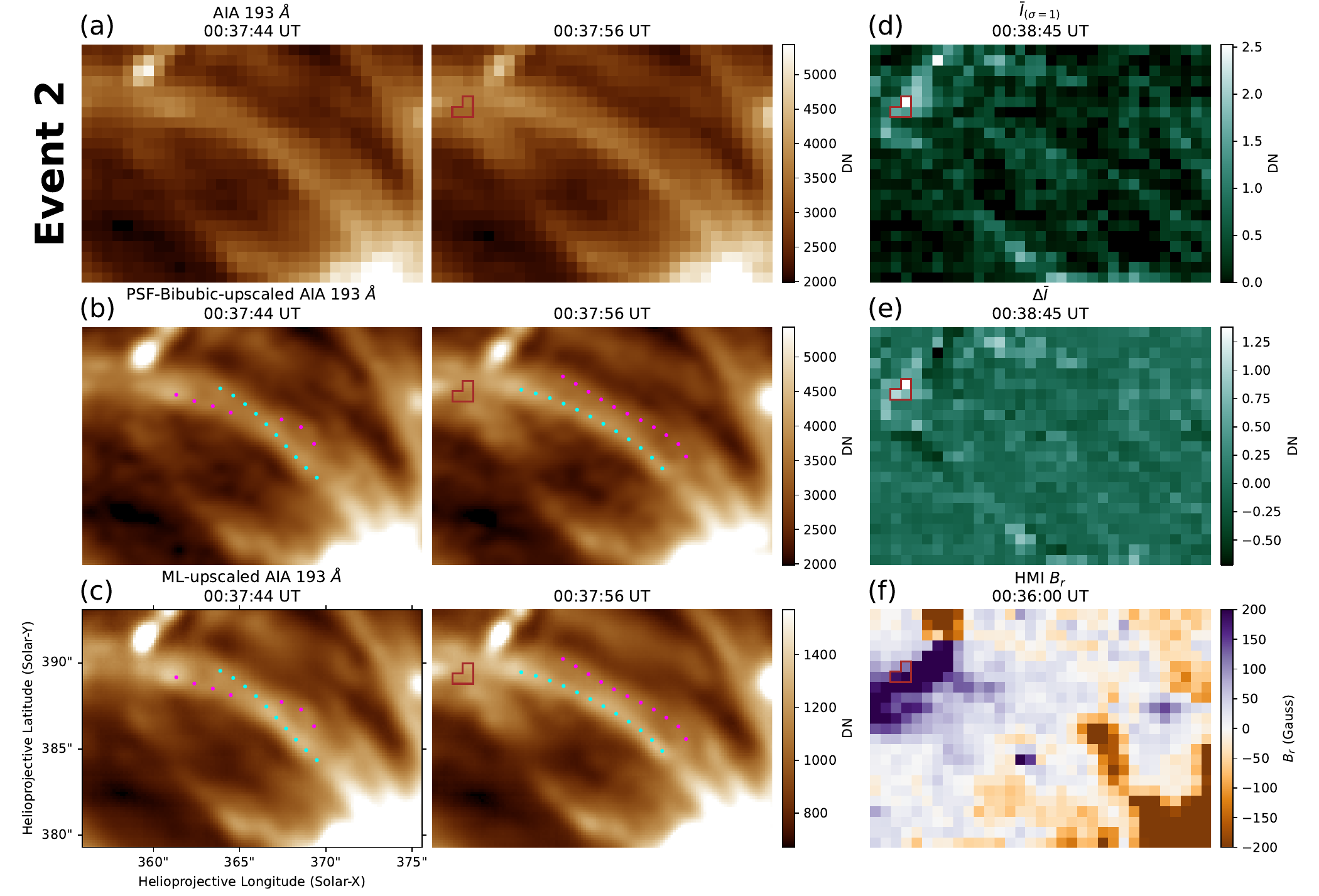}  
        \caption[]{\label{f4}   
  The evolution of the braiding strands on  22 June 2010 (Event 2).
{\em Panel a:\/} SDO/AIA 193 \AA\ images.  
{\em Panels b-c:\/} the Bicubic upscaled versions of the   PSF-matched AIA 193 \AA\ images 
 and ML-upscaled versions of the AIA 193 \AA\ images, on which  the various colored dot-lines indicate the various recognizable  strands.
{\em Panels d-e:\/}The images of $\bar{I}_{(\sigma=1)}$ and  $\Delta \bar{I}=\bar{I}_{(\sigma=1)}-\bar{I}_{(\sigma=4)}$, where the $\bar{I}_{(\sigma=1)}$ and $\bar{I}_{(\sigma=4)}$ is the Fe XVIII intensity of  $I(94 \AA\ )-I(211 \AA\ )/120-I(171 \AA\ )/450$ smoothed with a Gaussian kernel of 1 and 4 minutes, respectively.
  These  images  are taken at the peak time of the hot emission  indicated by the brown box in (d), which is only one  hot emission  identified from  00:35:08 UT to 00:45:08 UT. The brown box is also plotted in the other panels.
{\em Panel f:\/} The vertical component  ($B_{r}$) of the photospheric  vector data from SDO/HMI.      All images have the same FOV. 
 The evolution of the braiding strands is shown in a movie (anim4.mpeg) available online.
  }
  \end{figure*}  
    
  \begin{figure*}[hbtp]
  \centering
    \includegraphics[width=1.3\columnwidth]{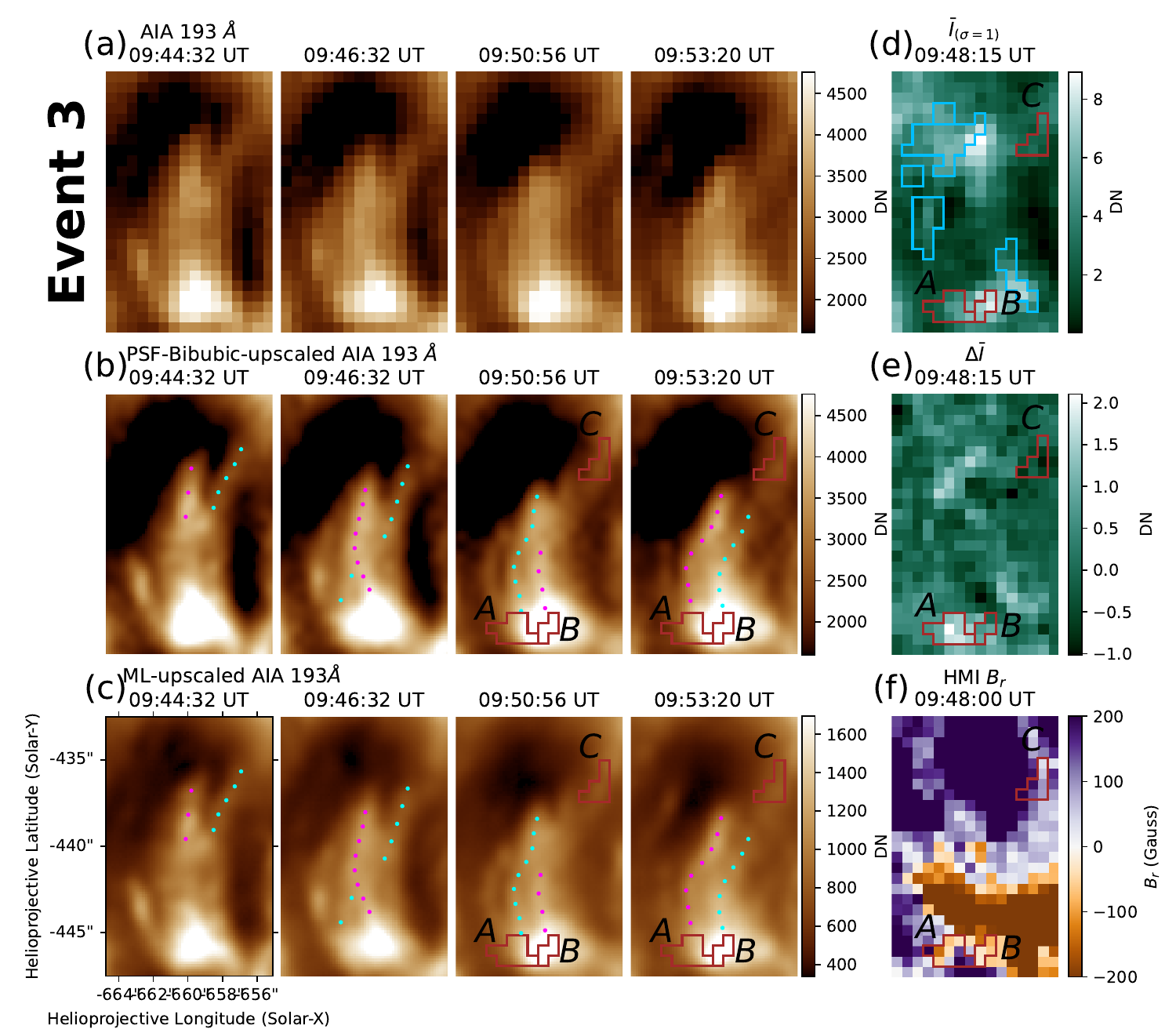} 
  \caption[]{\label{f5}   
  The evolution of the braiding strands on  30 August 2011 (Event 3).
 {\em Panel a:\/} SDO/AIA 193 \AA\ images.  
{\em Panels b-c:\/} the Bicubic upscaled versions of the   PSF-matched AIA 193 \AA\ images 
 and ML-upscaled versions of the AIA 193 \AA\ images, on which  the various colored dot-lines indicate the various recognizable  strands.
{\em Panels d-e:\/} The images of $\bar{I}_{(\sigma=1)}$ and  $\Delta \bar{I}=\bar{I}_{(\sigma=1)}-\bar{I}_{(\sigma=4)}$, where the $\bar{I}_{(\sigma=1)}$ and $\bar{I}_{(\sigma=4)}$ is the Fe XVIII intensity of  $I(94 \AA\ )-I(211 \AA\ )/120-I(171 \AA\ )/450$ smoothed with a Gaussian kernel of 1 and 4 minutes, respectively.
The boxes in {\em d} outline   the locations of all  hot emissions that are  identified from 09:44:08 UT to 09:54:08 UT, and the three brown boxes marking the hot emissions A, B, and C are also  plotted on the other panels.
 These  images  in (d) and (e) are taken at the peak time of the hot emission A. 
 {\em Panel f:\/}  The vertical component  ($B_{r}$) of the photospheric  vector data from SDO/HMI.      All images have the same FOV. 
 The evolution of the braiding strands is shown in a movie (anim5.mpeg) available online.
  }
  \end{figure*}  
  
   \section{Results} \label{s3}

   \begin{figure*}
  \centering
     \includegraphics[width=1.9\columnwidth]{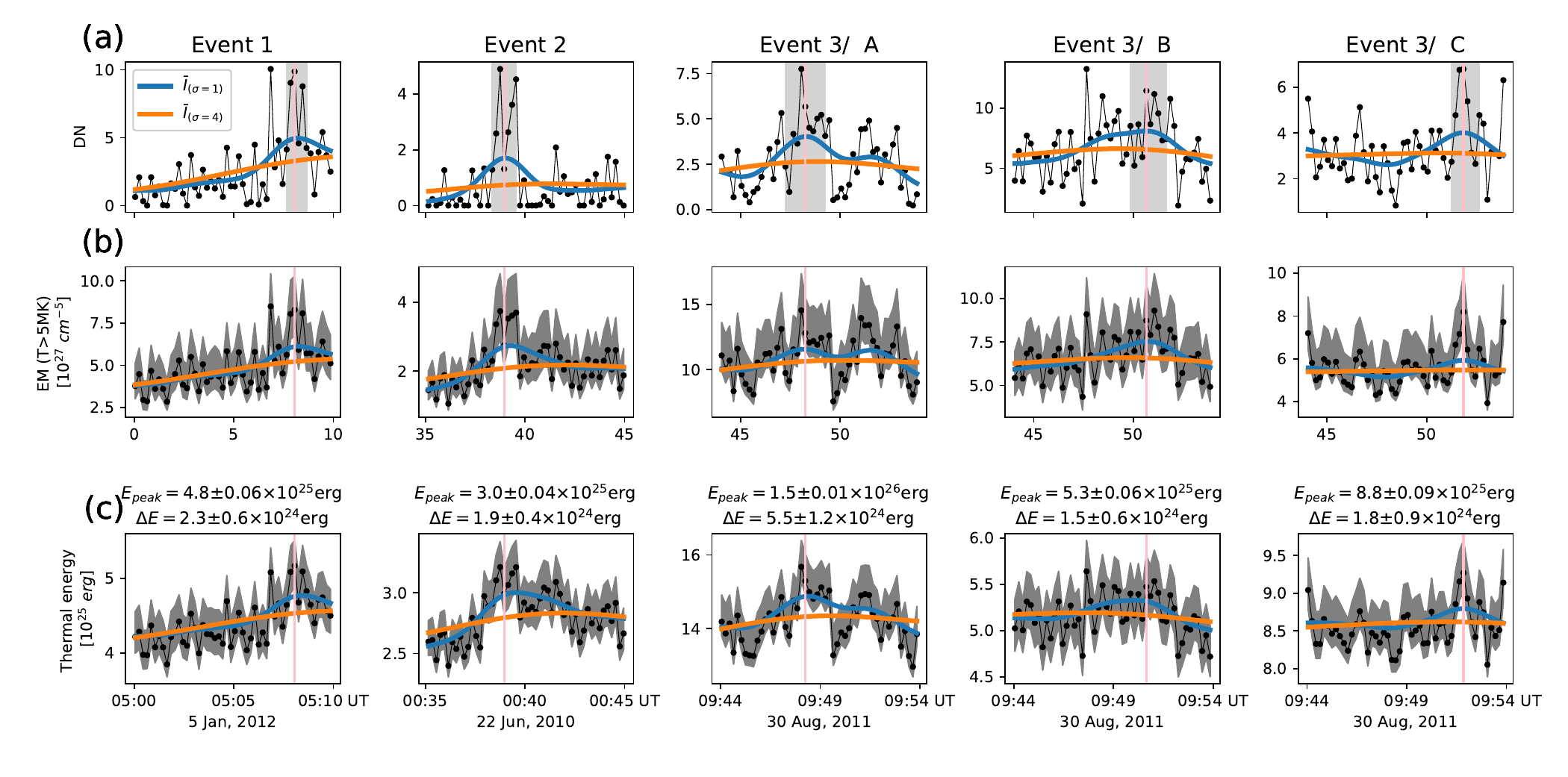} 
  \caption[]{\label{f6}   
  The evolutions of the identified hot emissions. 
{\em Panel a:\/} The light curves of  the  Fe XVIII intensity of  $I(94 \AA\ )-I(211 \AA\ )/120-I(171 \AA\ )/450$ from the hot emissions outlined by the brown boxes in Figs.  \ref{f3}, \ref{f4}, and \ref{f5}. In each panel, the light-gray area indicates the time period in which the value of  $\Delta \bar{I}$ is higher than 1.5 times the standard deviation of $\bar{I}_{\sigma =1}$.
{\em Panel b-c:\/} The  evolutions of EM at 5~-~10 MK and thermal energy $E$ from the locations of the hot emissions, in which the gray shaded area denotes the standard deviation
of 200 Monte Carlo simulations by adding random AIA instrument noise into the EM inversion.
In each panel, the dots with thin lines indicate the raw data;   the blue and orange curves indicate the   data smoothed with a Gaussian kernel of  1 and 4 minutes, respectively.
The vertical line indicates the peak time of $\Delta \bar{I}$, which amounts to $\bar{I}_{(\sigma=1)}-\bar{I}_{(\sigma=4)}$.
  }
  \end{figure*}  
    
  To investigate  the nanoflare candidates,  this work focuses on the braiding strands recognized in the  upscaled AIA 193 \AA\ images. We investigated the AIA 193 \AA\ observation of a sample of non-flare active regions (\citeads{Schmelz}), all of which were not covered by the  Hi-C observation.
We present three events as follows.

\subsection{Overview of the recognizable braiding strands and their evolutions}\label{s3.1}

   In Event 1, a  bifurcated loop is observed on the raw AIA 193 \AA\ images (Fig. \ref{f3}a). Subsequently, the loop seemingly departed into two separate ones.
   Both the PFS-Bicubic-upscaled and  ML-upscaled images show more detail of the evolution of the loop.
 In these upscaled   images (Figs.  \ref{f3}b and \ref{f3}c), the AIA loop appears to be 
  a pair of strands braiding with each other at the beginning of Event 1. 
The disappearance  of the thinner strand  (indicated by the magenta dot-line at 05:04:56 UT) was  followed by the appearance of a newly-formed strand (indicated by the magenta dot-line at 05:09:20 UT), which was parallel with the thicker one that seem to be nearly unchanged (indicated by the cyan dot-lines).

 Figs.  \ref{f4} and  \ref{f5} show  another two examples (Events 2 and 3) exhibiting  similar evolutions of  substructures from braiding to parallel with each other,  which were observed on  both the PFS-Bicubic-upscaled  and the ML-upscaled AIA 193 \AA\ images but were hardly seen in the raw AIA images.

   \subsection{  Transient brightenings   at the footpoints of the braiding structures}
  We use an empirical approach to isolate the hot plasma component (produced by  Fe XVIII emission) present in the 94 channel for coronal diagnostics of impulsive heating. A reasonable estimate of Fe XVIII emission is $I(94 \AA\ )-I(211 \AA\ )/120-I(171 \AA\ )/450$ (\citeads{Zanna}). The process eliminates the warm plasma component from the emission observed in the 94 channel, which is around 1 MK. 
  The peak formation temperature of Fe XVIII is at 7.1 MK (\citeads{Warren12}), but  Fe XVIII emission from plasma at 3 MK may detected due to the large fraction of plasma present at this temperature  (\citeads{Zanna}). 
  This information shows that the hot emissions are typically believed to have a temperature of at least 3 MK.
  To improve its signal-to-noise ratio, the light curve of the Fe XVIII intensity from each AIA pixel is smoothed in time, and the resulting $\bar{I}_{(\sigma =1)}$ and $\bar{I}_{(\sigma =4)}$ indicates   the  data
 smoothed  with a  Gaussian kernel of 1 and  4 minutes.
We apply the high-pass filter of the Fe XVIII intensity, corresponding to $\bar{I}_{(\sigma =1)}- \bar{I}_{(\sigma =4)}$, to extract the short hot emissions since the durations of the fluctuations in the 94 \AA\ channel in response to the nanoflare-scale heat pulses last for minutes (\citeads{Reale11,Tajfirouze}).
A pixel-wise emission enhancement   is identified when  $\Delta \bar{I}$ is  greater  than 1.5 times the standard deviation of $\bar{I}_{(\sigma =1)}$ during at least one minute. The decision to use a threshold of 1.5 standard deviations was made because picking a standard deviation with a smaller multiple will likely pick up some trivial disturbances. When such three or more adjacent  pixel-wise emission enhancements are detected at the same time, we define them as a transient  
 hot  emission.

 The hot emissions are found in all  events studied here.
As shown in Figs.  \ref{f3}d, \ref{f4}d, and \ref{f5}d, the brown boxes outline the concerned hot emissions, which seem to be associated with the evolutions of the braiding structures.
Specifically, a  hot emission took place at one of the footpoints of the braiding stands in Events  1 and 2 (Figs.  \ref{f3} and \ref{f4}), respectively;
 three hot emissions, in turn, occurred in the southeast, southwest, and northwest endpoints of the evolving strands in Event 3 (Fig. \ref{f5} and its accompanying  animation). 

 Fig. \ref{f6}a presents each light curve of the   Fe XVIII intensity from each hot emission.  The light-gray areas show that the fluctuations in  Fe XVIII intensity last for 1.2-2.2 minutes.
  Each peak time indicated by the vertical line corresponds to the moment when  $\Delta \bar{I}$ reaches its maximum.
  The comparisons of the peak times and  evolutions of the braiding strands (Figs.  \ref{f3}, \ref{f4}, \ref{f5}, and their accompanying animations) reveal that the hot emissions always reached their peaks  before the braiding strands evolved into two parallel ones.
 Such hot plasma provides evidence supporting that the energy was released from the magnetic reconnection (\citeads{Klimchuk,Schmelz}) during the morphological changes of the braiding strands. 

 Since the plasma along a given line-of-sight may have a range of temperatures rather than being isothermal, it is common to describe the coronal temperature distribution by  reconstruction of differential emission measures (DEMs).
 Here, we apply a Fast, Simple, Robust algorithm (\citeads{Plowman}) to inverse the DEM from AIA images in the six optically thin wavelengths, including 94 \AA\ , 131 \AA\ , 171 \AA\ , 193 \AA\ , 211 \AA\ , and 335 \AA\ .
 The temperature points in the inversion are selected to range from $10^{5.5}$   to $10^{7.0}$ K.
 The choice of  the maximum temperature of 10 MK ensures that  the hot emission would be not overestimated, since  the emission above 10 MK is  less well constrained by the six AIA channels.
The DEM analysis demonstrates that the hot plasmas include a significant high-temperature component ($T>5 MK$), as shown in Fig. \ref{f6}b., which presents the evolutions of emission measure (EM) at 5–10 MK from each hot emission.

Since the  coronal plasma is multi-thermal, we estimate the thermal energy as Equation 12 in \citetads{Aschwanden15}, where the volume of the transient emission is estimated as $V=A^{3/2}$, where $A$ refers to the area of each brightening, and unity filling factor is assumed.
  In Fig. \ref{f6}c, the time evolution of the thermal energy of each hot emission shows that $E_{peak}$  ranges  from   $3.0 \times 10^{25}$ to $1.5 \times 10^{26} ~erg$  and $\Delta E_{peak}$   ranges   $1.5\times 10^{24} -5.5 \times 10^{24} ~erg$.
    Here, the values of $E$  and  $\Delta E$ are also estimated from the amount of energy   smoothed     with a  Gaussian kernel of 1 minute (blue curves in Fig. \ref{f6}c) and  its difference with respect to  that   smoothed  with a  Gaussian kernel of 4 minutes (orange curves in Fig. \ref{f6}c).
  This amount of change in the thermal energy $\Delta E_{peak}$  corresponds to the level of the most common nanoflare energy suggested by \citetads{Parker88}.

 The hot emissions were often found to be rooted  in the unipolar region according to  simultaneous measurements of the radial component   $B_{r}$ of the photospheric magnetic field from SDO/HMI, such as in Events 1 and 2 (Figs.  \ref{f3}f and \ref{f4}f).
This excludes the possibility that the magnetic reconnections between opposite-polarity magnetic flux on the photosphere produce the hot plasma at the  endpoints of the loops (\citeads{Samanta}) unless there is minority-polarity invisible in the HMI magnetogram (\citeads{Wang}).
When compared to Events 1 and 2, which was centered at ($-190{\arcsec}$, $280{\arcsec}$) and ($370{\arcsec}$, $385{\arcsec}$)  from Sun center, respectively, Event 3  (Fig. \ref{f5}f) was centered at ($-660{\arcsec}$, $-440{\arcsec}$), and was then observed farther from the disk center. Determining the magnetic polarities in Event 3 from the HMI is therefore challenging due to near-limb projection effects.

   \subsection{Coronal Magnetic Extrapolation}  
      \begin{figure}[h!]
  \centering
  \includegraphics[width=1\columnwidth]{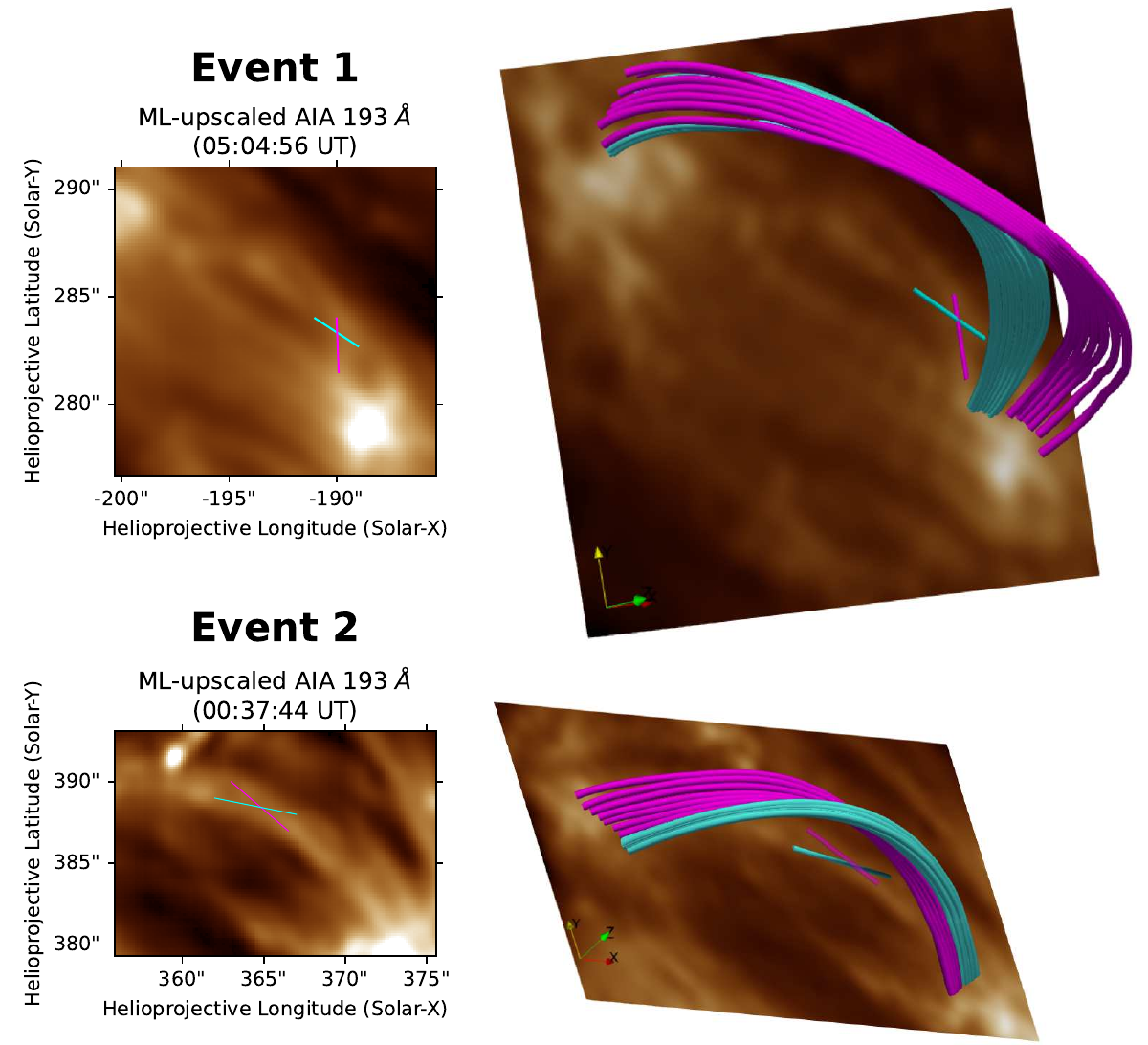} 
  \caption[]{\label{f7}   
  The modeled magnetic field lines matching the loops.
  {\em Left column:\/} The ML-upscaled versions of AIA 193 \AA\ images taken in Events 1 ({\em top}) and 2 ({\em bottom}), in which   the magenta and cyan lines outline the two loop stands.
The misalignment angle $\theta$ between the two lines  amounts to $55^{\circ}$ and  $29^{\circ}$ in Events 1 and 2, respectively. 
{\em Right column:\/} deposited upon the ML-upscaled AIA images, the field lines are traced from the NLFFF field.
In Events 1 ({\em top}) and 2 ({\em bottom}), the NLFFF field is extrapolated based on the set of HMI vector data taken at 05:00:00 UT on  5 January 2012 and 00:36:00 UT on  22 June 2010, respectively. 
  }
  \end{figure}

    The braiding strands may mark the bundles of coronal magnetic flux winding about each other (\citeads{Berger}).
    This is  supported by the comparisons of the braiding structures and the  Nonlinear Force-free Field (NLFFF) coronal magnetic fields, which are constructed   by the optimization method (\citeads{Wheatland}) with the required boundary conditions being provided by HMI vector data.
  The modeled field  shows  good alignment to the coronal loops in Events 1 and 2, suggesting that  force-free extrapolation could be considered as a consistent model of the  corona in these events (\citeads{Rosa}).
    Similar to the appearance of the  braiding strands, the modeled field aligning well with the AIA loop consists of   two bundles of modeled field lines twisting with each other (Figs.  \ref{f7}).
    Again, possibly due to magnetogram degradation by near-limb projection effects, the NLFFF field even failed to match the majority of coronal loops in Event 3.  
    
   The crossing manners of the strands imply that a localized  tangential discontinuity (\citeads{Parker87}) of the magnetic field may exist at the crossing site of the  braiding strands.
   The magnetic free energy carried in the localized current sheet would convert into heat and kinetic energy when magnetic reconnection occurs there.
   The amount of free energy is of order $B^{2}_{\bot}V/8\pi$, where   $B_{\bot}$ is of the order of $B sin(\theta)$.
   Here,  $B$ is estimated as magnetic field strength of the NLFFF lines, which amounts to  $\sim$ 60G and  $\sim$ 105G in Event 1 and 2, respectively.
The  discontinuity in the field direction $\theta$ is assumed to be of the order of    the misalignment angle $\theta$ between the two  braiding strands, which 
 is about $55^\circ$ and $29^\circ$   in Event 1 and 2, respectively.
Moreover,  we assume that  characteristic length $\Delta L$ is  of the order of  1{\arcsec} , corresponding to the characteristic width of the  strands detected in the upscaled AIA 193 \AA\ images (see  Appendix \ref{append:b}), and then 
the Volume $V \approx (\Delta L)^3$.
 With the number estimated above the magnetic free energy associated with a discontinuity is of the order of $3.7 \times 10^{25} erg$ and $3.9 \times 10^{25} erg$ for Event 1 and Event 2, respectively.
 Therefore, the amount of magnetic free energy could account for the thermal energy of order of $10^{24}~erg$ released by  a nanoflare.

\section{Discussions}\label{s4}

      \begin{figure}
  \centering
  \includegraphics[width=1\columnwidth]{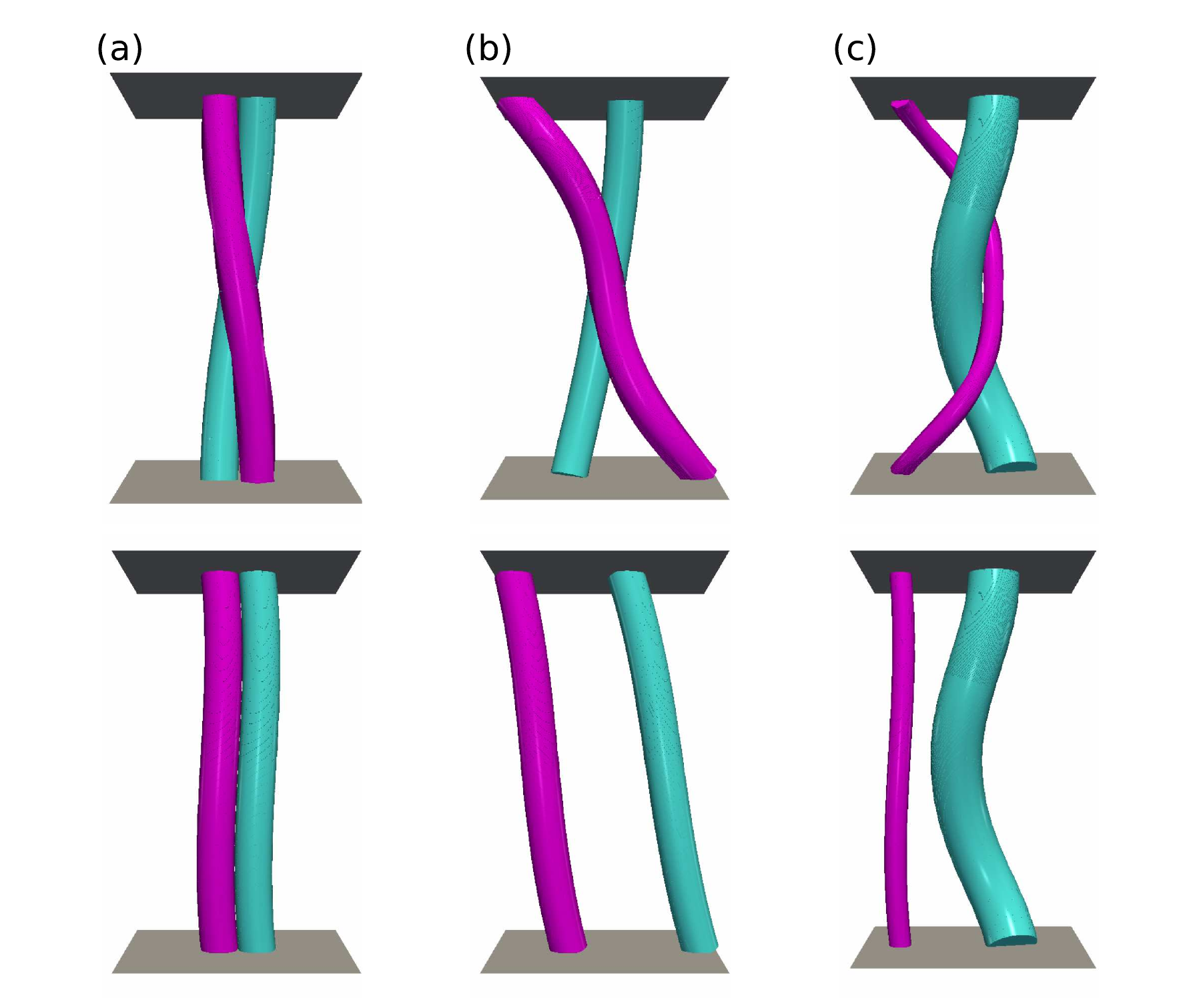} 
  \caption[]{\label{f8}   
   Picture of evolutions of two flux tubes with various braid pattens.
 {\em Top row:\/} The flux tubes braiding with various patterns.
 {\em bottom row:\/}  Their  final states after occurrences  of magnetic reconnection at their crossings.  
{\em Panel a:\/} The two coherent flux tubes braid  with each other in a well-combed pattern.
{\em Panel b-d:\/} The two flux tubes  braid  with their legs separating from each other randomly. 
The two flux tubes in {\em panels a} and {\em b} initially have identical flux and,  while the two tubes  in {\em panels c} and {\em d} have nonidentical flux and tubes colored cyan have more axial flux than that colored magenta.
 The pairs of flux tubes in {\em panels a-c} initially braid with one crossing and the  two tubes in {\em panel d} braid with two crossings.
  }
  \end{figure}

The investigation of the widths of the detectable loop structures (see Appendix \ref{append:b}) shows that the  widths of strands recognized in  both the PSF-bicubic upscaled and  ML-upscaled AIA 193 \AA\  images  range from  0{\arcsec}.7 to 1{\arcsec}.3  (Fig. \ref{f10}),  corresponding to a physical size ranging from about   500 $Km$ to 1000 $Km$,
and that each  pair of strands is resolved in an  AIA loop with characteristic width of  2{\arcsec} to  3{\arcsec} (\citeads{Aschwanden17}).  
The uncertainty of width of the upscaled structures amounts to $ \pm 0{\arcsec}.3$  and is roughly determined by the results from various networks trained with slightly different training sets.
Therefore,  the  upscaled AIA 193 \AA\  images have a performance to resolve  the strands with widths less than 1{\arcsec}.2, which could hardly be resolved on the raw AIA images with a  spatial resolution of  1{\arcsec}.2.
According to \citetads{Brooks},   the lowest and mean Gaussian widths of loops observed in  Hi-C was about 90 km and 270 km, or 0{\arcsec}.12 and 0{\arcsec}.37, respectively. By contrast, strands in the upscaled AIA 193 \AA\ images always have  widths  no less than a pixel size of 0{\arcsec}.6 of AIA images. This is mostly because not more information can be extracted from the AIA images and then the  minimum width of the strands detected on the ML-upscaled images is close to  the pixel size of AIA images.

  From our observations, we infer the schematic picture of Fig. \ref{f8} for  two flux tubes braiding  in various patterns and their subsequent evolutions due to magnetic reconnection.
  In the upscaled AIA 193 \AA\ image we could recognize some pairs of strands with their footpoints separated randomly (e.g., Fig. \ref{f8}b-c), instead of  a pair of strands braiding coherently in a well-combed pattern (Fig. \ref{f8}a), which is difficult to observe due to the  limit of the resolution and the effect of  cross-field diffusion electrons in tangled magnetic fields (\citeads{Galloway,Berger}).
  Fig. \ref{f8}b shows  two  identical tubes braiding with each other and the magnetic reconnection between them would result in a total exchange of their footpoints.
    The picture could explain the morphological evolution exhibited in Events 2 and 3 (Figs.  \ref{f4} and \ref{f5}), in which the two braiding strands evolved to be two parallel ones.
    Figs.  \ref{f8}c  shows the evolutions of two braiding flux tubes with nonequal axial flux.
     As  the two nonidentical flux tubes  collide with each other,  the reconnection halts once all  flux of the smaller tube has reconnected, and a final state is that only the outer shell of the larger flux tube is reconnected and 
    the rest is unreconnected (\citeads{Linton}). 
 Moreover, when the  two nonidentical  flux tubes braid with two crossings (Fig. \ref{f8}d), magnetic reconnection between them would occur twice and then produce two thoroughly separated tubes.
        Consistently, we can see in Event 1 (Fig. \ref{f3}) that the unequal width strands braided with two apparent crossings and that subsequently, the thinner one became parallel with the other one nearly unchanged.
Accordingly, the morphological evolutions of the two braiding strands presented here are  consistent with the schematic pictures of two braiding bundles of magnetic flux driven to reconnect with each other.

\citetads{Pontin} claimed that the existence of crossed loop strands does not always imply magnetic discontinuities and subsequent magnetic reconnection. Here, further evidence for the occurrence of magnetic reconnection between the braiding strands is provided  by the transient hot emission around the footpoints of the loop strands. The impulsive brightenings at the footpoints of the loops were also noted to happen at lower temperatures between 1 and 2 MK (\citeads{Regnier14,Subramanian}). By contrast,  transient emissions with higher temperatures were detected in  the  events reported here. According to the EM analysis, the hot emissions contain a temperature component greater than 5 MK. This value can be easily obtained using reconnection models (\citeads{Schmelz}) but are difficult to get using wave models (\citeads{Van17}).

 The footpoints of the loops could have been heated by the energetic electrons produced by  magnetic reconnections high in the corona, even though  hot plasma components were undetectable   at the locations of reconnection sites (e.g. \citeads{Zhang19}).   
This is due to the fact that there were low densities higher in the corona and that, as predicted by a simulation conducted by \citetads{Polito}, the energtic electrons then deposited raw kinetic energy in the corona until they reached the low corona with an increasing density.
  Given that loops are isothermal along their coronal parts,  heat pulses at their footpoints might cause the loop to heat up and become  denser as a result of thermal conduction and chromospheric evaporation. However, the measurable  increase in the 94 \AA\ emission was also absent throughout the entirety of the loops reported here. This could be as a result of smaller 94 \AA\ variations in the loops' lower-density sections than at their footpoints ( \citeads{Tajfirouze16}).  
  To explore the locations and temperatures of hot plasmas created in response to the braided magnetic field forming the coronal loops and to evaluate the implications provided from this study, more simulation model findings would be helpful.
 
\section{ Summary} \label{s5}

 In this article, the evolutions of the strands braiding with each other  in the apparent single AIA loops are presented. 
  The main results are summarized as follows:
  
  \begin{itemize}
\item[1.]  We performed two validations to confirm that the substructures recorded in the AIA pictures could be seen after the images were appropriately enhanced and upscaled. The width of the strands that make up what appear to be single AIA loops  ranges from   0{\arcsec}.7 to 1{\arcsec}.3 
\item[2.] The braided substructured loops were discovered to closely match the twisted NLFFF field lines in some events that are observed near the disk-center and then the photospheric magnetic data is properly recorded. 
The magnetic free energy in the modeled field is sufficient to match the thermal energy required for a nanoflare.
\item[3.] The braided strands developed into pairs of almost parallel ones together with the hot emission that was present at their footpoints,  supporting the occurrence of magnetic reconnection in the coronal loops shown in the AIA images.
\end{itemize}

A comparison of the raw AIA, ML-upscaled AIA, and Hi-C images (Figs. \ref{f2} and \ref{f9}) reveals that the ML-algorithm only detects a portion of the features at a scale smaller than $\sim$ 1{\arcsec}.2 that can be seen by Hi-C.   As a result, it would appear that the evolutions of the braided structures pertaining to nanoflares are predicted to be discovered more frequently in the higher resolution observations, such EUI on board Solar Orbiter. 

\begin{acknowledgements} 
     The authors are grateful to the anonymous referee for detailed comments and useful suggestions that improved this manuscript.
We acknowledge the High resolution Coronal Imager instrument team for making the flight data publicly available. MSFC/NASA led the mission and partners include the Smithsonian Astrophysical Observatory in Cambridge, Mass.; Lockheed Martin's Solar Astrophysical Laboratory in Palo Alto, Calif.; the University of Central Lancashire in Lancashire, England; and the Lebedev Physical Institute of the Russian Academy of Sciences in Moscow.
 The NASA/SDO data used here are courtesy of the AIA  and HMI science teams.
  This work is supported  by the National Key Research and Development Program of China (2019YFA0405000),
and  the Natural Science Foundation of China under grants 12273106, 12073077, 12163004,   U2031140, 12073072, 11933009, 12273108, 12203097, 12003068, 11873088, 12173084,  and  11973088, the CAS "Light of West China" Program, and the Strategic Priority Research Program of Chinese Academy of Sciences, Grant No. XDB 41000000.

\end{acknowledgements}

\bibliographystyle{aa-note} 
\bibliography{bi_v4}      

\begin{thebibliography}{70}
\expandafter\ifx\csname natexlab\endcsname\relax\def\natexlab#1{#1}\fi

\bibitem[{{Antolin} {et~al.}(2021){Antolin}, {Pagano}, {Testa}, {Petralia}, \&
  {Reale}}]{Antolin}
{Antolin}, P., {Pagano}, P., {Testa}, P., {Petralia}, A., \& {Reale}, F. 2021,
  Nature Astronomy, 5, 54 \csname Antolinlink\endcsname~\csname
  Antolinnote\endcsname

\bibitem[{{Aschwanden} {et~al.}(2015){Aschwanden}, {Boerner}, {Ryan}, {Caspi},
  {McTiernan}, \& {Warren}}]{Aschwanden15}
{Aschwanden}, M.~J., {Boerner}, P., {Ryan}, D., {et~al.} 2015, \apj, 802, 53
  \csname Aschwanden15link\endcsname~\csname Aschwanden15note\endcsname

\bibitem[{{Aschwanden} \& {Peter}(2017)}]{Aschwanden17}
{Aschwanden}, M.~J. \& {Peter}, H. 2017, \apj, 840, 4 \csname
  Aschwanden17link\endcsname~\csname Aschwanden17note\endcsname

\bibitem[{{Asgari-Targhi} {et~al.}(2019){Asgari-Targhi}, {van Ballegooijen}, \&
  {Davey}}]{Asgari19}
{Asgari-Targhi}, M., {van Ballegooijen}, A.~A., \& {Davey}, A.~R. 2019, \apj,
  881, 107 \csname Asgari19link\endcsname~\csname Asgari19note\endcsname

\bibitem[{Aydin {et~al.}(2008)Aydin, Mantiuk, Myszkowski, \& Seidel}]{Aydin}
Aydin, T.~O., Mantiuk, R., Myszkowski, K., \& Seidel, H.-P. 2008, ACM
  Transactions on Graphics, 27, 69 \csname Aydinlink\endcsname~\csname
  Aydinnote\endcsname

\bibitem[{{Bai} {et~al.}(2021){Bai}, {Liu}, {Deng}, {Jiang}, {Guo}, {Bi},
  {Feng}, {Jin}, {Cao}, {Su}, \& {Ji}}]{Bai}
{Bai}, X., {Liu}, H., {Deng}, Y., {et~al.} 2021, \aap, 652, A143 \csname
  Bailink\endcsname~\csname Bainote\endcsname

\bibitem[{{Berger} \& {Asgari-Targhi}(2009)}]{Berger}
{Berger}, M.~A. \& {Asgari-Targhi}, M. 2009, \apj, 705, 347 \csname
  Bergerlink\endcsname~\csname Bergernote\endcsname

\bibitem[{{Bi} {et~al.}(2020){Bi}, {Yang}, {Li}, {Dong}, \& {Ji}}]{Bi}
{Bi}, Y., {Yang}, B., {Li}, T., {Dong}, Y., \& {Ji}, K. 2020, \apjl, 891, L40
  \csname Bilink\endcsname~\csname Binote\endcsname

\bibitem[{{Boerner} {et~al.}(2012){Boerner}, {Edwards}, {Lemen}, {Rausch},
  {Schrijver}, {Shine}, {Shing}, {Stern}, {Tarbell}, {Title}, {Wolfson},
  {Soufli}, {Spiller}, {Gullikson}, {McKenzie}, {Windt}, {Golub}, {Podgorski},
  {Testa}, \& {Weber}}]{Boerner}
{Boerner}, P., {Edwards}, C., {Lemen}, J., {et~al.} 2012, \solphys, 275, 41
  \csname Boernerlink\endcsname~\csname Boernernote\endcsname

\bibitem[{{Brooks} {et~al.}(2013){Brooks}, {Warren}, {Ugarte-Urra}, \&
  {Winebarger}}]{Brooks}
{Brooks}, D.~H., {Warren}, H.~P., {Ugarte-Urra}, I., \& {Winebarger}, A.~R.
  2013, \apjl, 772, L19 \csname Brookslink\endcsname~\csname
  Brooksnote\endcsname

\bibitem[{{Chitta} {et~al.}(2022){Chitta}, {Peter}, {Parenti}, {Berghmans},
  {Auch{\`e}re}, {Solanki}, {Aznar Cuadrado}, {Sch{\"u}hle}, {Teriaca},
  {Mandal}, {Barczynski}, {Buchlin}, {Harra}, {Kraaikamp}, {Long}, {Rodriguez},
  {Schwanitz}, {Smith}, {Verbeeck}, {Zhukov}, {Liu}, \& {Cheung}}]{Chitta22}
{Chitta}, L.~P., {Peter}, H., {Parenti}, S., {et~al.} 2022, \aap, 667, A166
  \csname Chitta22link\endcsname~\csname Chitta22note\endcsname

\bibitem[{{Chitta} {et~al.}(2020){Chitta}, {Peter}, {Priest}, \&
  {Solanki}}]{Chitta20}
{Chitta}, L.~P., {Peter}, H., {Priest}, E.~R., \& {Solanki}, S.~K. 2020, \aap,
  644, A130 \csname Chitta20link\endcsname~\csname Chitta20note\endcsname

\bibitem[{{Cirtain} {et~al.}(2013){Cirtain}, {Golub}, {Winebarger}, {de
  Pontieu}, {Kobayashi}, {Moore}, {Walsh}, {Korreck}, {Weber}, {McCauley},
  {Title}, {Kuzin}, \& {Deforest}}]{Cirtain}
{Cirtain}, J.~W., {Golub}, L., {Winebarger}, A.~R., {et~al.} 2013, \nat, 493,
  501 \csname Cirtainlink\endcsname~\csname Cirtainnote\endcsname

\bibitem[{{De Pontieu} {et~al.}(2007){De Pontieu}, {McIntosh}, {Carlsson},
  {Hansteen}, {Tarbell}, {Schrijver}, {Title}, {Shine}, {Tsuneta}, {Katsukawa},
  {Ichimoto}, {Suematsu}, {Shimizu}, \& {Nagata}}]{DePontieu}
{De Pontieu}, B., {McIntosh}, S.~W., {Carlsson}, M., {et~al.} 2007, Science,
  318, 1574 \csname DePontieulink\endcsname~\csname DePontieunote\endcsname

\bibitem[{{De Rosa} {et~al.}(2009){De Rosa}, {Schrijver}, {Barnes}, {Leka},
  {Lites}, {Aschwanden}, {Amari}, {Canou}, {McTiernan}, {R{\'e}gnier},
  {Thalmann}, {Valori}, {Wheatland}, {Wiegelmann}, {Cheung}, {Conlon},
  {Fuhrmann}, {Inhester}, \& {Tadesse}}]{Rosa}
{De Rosa}, M.~L., {Schrijver}, C.~J., {Barnes}, G., {et~al.} 2009, \apj, 696,
  1780 \csname Rosalink\endcsname~\csname Rosanote\endcsname

\bibitem[{{Del Zanna}(2013)}]{Zanna}
{Del Zanna}, G. 2013, \aap, 558, A73 \csname Zannalink\endcsname~\csname
  Zannanote\endcsname

\bibitem[{{D{\'\i}az Baso} \& {Asensio Ramos}(2018)}]{Dazbaso}
{D{\'\i}az Baso}, C.~J. \& {Asensio Ramos}, A. 2018, \aap, 614, A5 \csname
  Dazbasolink\endcsname~\csname Dazbasonote\endcsname

\bibitem[{Dong {et~al.}(2016)Dong, Loy, He, \& Tang}]{Dong}
Dong, C., Loy, C.~C., He, K., \& Tang, X. 2016, IEEE Transactions on Pattern
  Analysis and Machine Intelligence, 38, 295 \csname Donglink\endcsname~\csname
  Dongnote\endcsname

\bibitem[{{Dos Santos} {et~al.}(2021){Dos Santos}, {Bose}, {Salvatelli},
  {Neuberg}, {Cheung}, {Janvier}, {Jin}, {Gal}, {Boerner}, \&
  {Baydin}}]{Dos_santos}
{Dos Santos}, L. F.~G., {Bose}, S., {Salvatelli}, V., {et~al.} 2021, \aap, 648,
  A53 \csname Dos_santoslink\endcsname~\csname Dos_santosnote\endcsname

\bibitem[{Fan {et~al.}(2018)Fan, Yu, \& Huang}]{Fan2018}
Fan, Y., Yu, J., \& Huang, T.~S. 2018, in Proceedings of the IEEE Conference on
  Computer Vision and Pattern Recognition Workshops, 2621--2624 \csname
  Fan2018link\endcsname~\csname Fan2018note\endcsname

\bibitem[{{Galloway} {et~al.}(2006){Galloway}, {Helander}, \&
  {MacKinnon}}]{Galloway}
{Galloway}, R.~K., {Helander}, P., \& {MacKinnon}, A.~L. 2006, \apj, 646, 615
  \csname Gallowaylink\endcsname~\csname Gallowaynote\endcsname

\bibitem[{{Gupta} {et~al.}(2018){Gupta}, {Sarkar}, \& {Tripathi}}]{Gupta}
{Gupta}, G.~R., {Sarkar}, A., \& {Tripathi}, D. 2018, \apj, 857, 137 \csname
  Guptalink\endcsname~\csname Guptanote\endcsname

\bibitem[{He {et~al.}(2016)He, Zhang, Ren, \& Sun}]{He}
He, K., Zhang, X., Ren, S., \& Sun, J. 2016, in Proceedings of the IEEE
  Conference on Computer Vision and Pattern Recognition (CVPR) (Los Alamitos,
  CA, USA: IEEE Computer Society), 770--778 \csname Helink\endcsname~\csname
  Henote\endcsname

\bibitem[{{Hong} {et~al.}(2021){Hong}, {Liu}, {Bi}, {Xu}, {Yang}, {Yang}, {Su},
  {Xia}, \& {Ji}}]{Hong}
{Hong}, J., {Liu}, H., {Bi}, Y., {et~al.} 2021, \apj, 915, 96 \csname
  Honglink\endcsname~\csname Hongnote\endcsname

\bibitem[{{Ishikawa} {et~al.}(2017){Ishikawa}, {Glesener}, {Krucker},
  {Christe}, {Buitrago-Casas}, {Narukage}, \& {Vievering}}]{Ishikawa}
{Ishikawa}, S.-n., {Glesener}, L., {Krucker}, S., {et~al.} 2017, Nature
  Astronomy, 1, 771 \csname Ishikawalink\endcsname~\csname
  Ishikawanote\endcsname

\bibitem[{{Kim} {et~al.}(2019){Kim}, {Park}, {Lee}, {Moon}, {Bae}, {Lim},
  {Jang}, {Kim}, {Cho}, {Choi}, \& {Cho}}]{Kim}
{Kim}, T., {Park}, E., {Lee}, H., {et~al.} 2019, Nature Astronomy, 3, 397
  \csname Kimlink\endcsname~\csname Kimnote\endcsname

\bibitem[{{Klimchuk}(2006)}]{Klimchuk}
{Klimchuk}, J.~A. 2006, \solphys, 234, 41 \csname
  Klimchuklink\endcsname~\csname Klimchuknote\endcsname

\bibitem[{{Klimchuk}(2015)}]{Klimchuk15}
{Klimchuk}, J.~A. 2015, Philosophical Transactions of the Royal Society of
  London Series A, 373, 20140256 \csname Klimchuk15link\endcsname~\csname
  Klimchuk15note\endcsname

\bibitem[{{Kobayashi} {et~al.}(2014){Kobayashi}, {Cirtain}, {Winebarger},
  {Korreck}, {Golub}, {Walsh}, {De Pontieu}, {DeForest}, {Title}, {Kuzin},
  {Savage}, {Beabout}, {Beabout}, {Podgorski}, {Caldwell}, {McCracken},
  {Ordway}, {Bergner}, {Gates}, {McKillop}, {Cheimets}, {Platt}, {Mitchell}, \&
  {Windt}}]{Kobayashi}
{Kobayashi}, K., {Cirtain}, J., {Winebarger}, A.~R., {et~al.} 2014, \solphys,
  289, 4393 \csname Kobayashilink\endcsname~\csname Kobayashinote\endcsname

\bibitem[{Ledig {et~al.}(2017)Ledig, Theis, Huszár, Caballero, Cunningham,
  Acosta, Aitken, Tejani, Totz, Wang, \& Shi}]{Ledig}
Ledig, C., Theis, L., Huszár, F., {et~al.} 2017, in 2017 IEEE Conference on
  Computer Vision and Pattern Recognition (CVPR), 105--114 \csname
  Lediglink\endcsname~\csname Ledignote\endcsname

\bibitem[{{Lemen} {et~al.}(2012){Lemen}, {Title}, {Akin}, {Boerner}, {Chou},
  {Drake}, {Duncan}, {Edwards}, {Friedlaender}, {Heyman}, {Hurlburt}, {Katz},
  {Kushner}, {Levay}, {Lindgren}, {Mathur}, {McFeaters}, {Mitchell}, {Rehse},
  {Schrijver}, {Springer}, {Stern}, {Tarbell}, {Wuelser}, {Wolfson}, {Yanari},
  {Bookbinder}, {Cheimets}, {Caldwell}, {Deluca}, {Gates}, {Golub}, {Park},
  {Podgorski}, {Bush}, {Scherrer}, {Gummin}, {Smith}, {Auker}, {Jerram},
  {Pool}, {Soufli}, {Windt}, {Beardsley}, {Clapp}, {Lang}, \&
  {Waltham}}]{Lemen}
{Lemen}, J.~R., {Title}, A.~M., {Akin}, D.~J., {et~al.} 2012, \solphys, 275, 17
  \csname Lemenlink\endcsname~\csname Lemennote\endcsname

\bibitem[{{Li} {et~al.}(2015){Li}, {Peter}, {Chen}, \& {Zhang}}]{Li}
{Li}, L.~P., {Peter}, H., {Chen}, F., \& {Zhang}, J. 2015, \aap, 583, A109
  \csname Lilink\endcsname~\csname Linote\endcsname

\bibitem[{{Linton}(2006)}]{Linton}
{Linton}, M.~G. 2006, Journal of Geophysical Research (Space Physics), 111,
  A12S09 \csname Lintonlink\endcsname~\csname Lintonnote\endcsname

\bibitem[{{McIntosh} {et~al.}(2011){McIntosh}, {de Pontieu}, {Carlsson},
  {Hansteen}, {Boerner}, \& {Goossens}}]{McIntosh}
{McIntosh}, S.~W., {de Pontieu}, B., {Carlsson}, M., {et~al.} 2011, \nat, 475,
  477 \csname McIntoshlink\endcsname~\csname McIntoshnote\endcsname

\bibitem[{{Parker}(1987)}]{Parker87}
{Parker}, E.~N. 1987, \apj, 318, 876 \csname Parker87link\endcsname~\csname
  Parker87note\endcsname

\bibitem[{{Parker}(1988)}]{Parker88}
{Parker}, E.~N. 1988, \apj, 330, 474 \csname Parker88link\endcsname~\csname
  Parker88note\endcsname

\bibitem[{{Pineci} {et~al.}(2021){Pineci}, {Sadowski}, {Gaidos}, \&
  {Sun}}]{Pineci}
{Pineci}, A., {Sadowski}, P., {Gaidos}, E., \& {Sun}, X. 2021, \apjl, 910, L25
  \csname Pinecilink\endcsname~\csname Pinecinote\endcsname

\bibitem[{{Plowman} \& {Caspi}(2020)}]{Plowman}
{Plowman}, J. \& {Caspi}, A. 2020, \apj, 905, 17 \csname
  Plowmanlink\endcsname~\csname Plowmannote\endcsname

\bibitem[{{Polito} {et~al.}(2018){Polito}, {Testa}, {Allred}, {De Pontieu},
  {Carlsson}, {Pereira}, {Go{\v{s}}i{\'c}}, \& {Reale}}]{Polito}
{Polito}, V., {Testa}, P., {Allred}, J., {et~al.} 2018, \apj, 856, 178 \csname
  Politolink\endcsname~\csname Politonote\endcsname

\bibitem[{{Pontin} {et~al.}(2017){Pontin}, {Janvier}, {Tiwari}, {Galsgaard},
  {Winebarger}, \& {Cirtain}}]{Pontin}
{Pontin}, D.~I., {Janvier}, M., {Tiwari}, S.~K., {et~al.} 2017, \apj, 837, 108
  \csname Pontinlink\endcsname~\csname Pontinnote\endcsname

\bibitem[{{Priest} {et~al.}(2018){Priest}, {Chitta}, \& {Syntelis}}]{Priest}
{Priest}, E.~R., {Chitta}, L.~P., \& {Syntelis}, P. 2018, \apjl, 862, L24
  \csname Priestlink\endcsname~\csname Priestnote\endcsname

\bibitem[{{Reale}(2014)}]{Reale14}
{Reale}, F. 2014, Living Reviews in Solar Physics, 11, 4 \csname
  Reale14link\endcsname~\csname Reale14note\endcsname

\bibitem[{{Reale} {et~al.}(2011){Reale}, {Guarrasi}, {Testa}, {DeLuca},
  {Peres}, \& {Golub}}]{Reale11}
{Reale}, F., {Guarrasi}, M., {Testa}, P., {et~al.} 2011, \apjl, 736, L16
  \csname Reale11link\endcsname~\csname Reale11note\endcsname

\bibitem[{{R{\'e}gnier} {et~al.}(2014){R{\'e}gnier}, {Alexander}, {Walsh},
  {Winebarger}, {Cirtain}, {Golub}, {Korreck}, {Mitchell}, {Platt}, {Weber},
  {De Pontieu}, {Title}, {Kobayashi}, {Kuzin}, \& {DeForest}}]{Regnier14}
{R{\'e}gnier}, S., {Alexander}, C.~E., {Walsh}, R.~W., {et~al.} 2014, \apj,
  784, 134 \csname Regnier14link\endcsname~\csname Regnier14note\endcsname

\bibitem[{Salimans \& Kingma(2016)}]{Salimans}
Salimans, T. \& Kingma, D.~P. 2016, in Proceedings of the 30th International
  Conference on Neural Information Processing Systems, NIPS'16 (Red Hook, NY,
  USA: Curran Associates Inc.), 901–909 \csname
  Salimanslink\endcsname~\csname Salimansnote\endcsname

\bibitem[{{Samanta} {et~al.}(2019){Samanta}, {Tian}, {Yurchyshyn}, {Peter},
  {Cao}, {Sterling}, {Erd{\'e}lyi}, {Ahn}, {Feng}, {Utz}, {Banerjee}, \&
  {Chen}}]{Samanta}
{Samanta}, T., {Tian}, H., {Yurchyshyn}, V., {et~al.} 2019, Science, 366, 890
  \csname Samantalink\endcsname~\csname Samantanote\endcsname

\bibitem[{{Schmelz} {et~al.}(2015){Schmelz}, {Asgari-Targhi}, {Christian},
  {Dhaliwal}, \& {Pathak}}]{Schmelz}
{Schmelz}, J.~T., {Asgari-Targhi}, M., {Christian}, G.~M., {Dhaliwal}, R.~S.,
  \& {Pathak}, S. 2015, \apj, 806, 232 \csname Schmelzlink\endcsname~\csname
  Schmelznote\endcsname

\bibitem[{{Schou} {et~al.}(2012){Schou}, {Scherrer}, {Bush}, {Wachter},
  {Couvidat}, {Rabello-Soares}, {Bogart}, {Hoeksema}, {Liu}, {Duvall}, {Akin},
  {Allard}, {Miles}, {Rairden}, {Shine}, {Tarbell}, {Title}, {Wolfson},
  {Elmore}, {Norton}, \& {Tomczyk}}]{Schou}
{Schou}, J., {Scherrer}, P.~H., {Bush}, R.~I., {et~al.} 2012, \solphys, 275,
  229 \csname Schoulink\endcsname~\csname Schounote\endcsname

\bibitem[{{Shen}(2021)}]{Shen}
{Shen}, Y. 2021, Proceedings of the Royal Society of London Series A, 477, 217
  \csname Shenlink\endcsname~\csname Shennote\endcsname

\bibitem[{{Shi} {et~al.}(2016){Shi}, {Caballero}, {Husz{\'a}r}, {Totz},
  {Aitken}, {Bishop}, {Rueckert}, \& {Wang}}]{Shi}
{Shi}, W., {Caballero}, J., {Husz{\'a}r}, F., {et~al.} 2016, arXiv e-prints,
  arXiv:1609.05158 \csname Shilink\endcsname~\csname Shinote\endcsname

\bibitem[{{Subramanian} {et~al.}(2018){Subramanian}, {Kashyap}, {Tripathi},
  {Madjarska}, \& {Doyle}}]{Subramanian}
{Subramanian}, S., {Kashyap}, V.~L., {Tripathi}, D., {Madjarska}, M.~S., \&
  {Doyle}, J.~G. 2018, \aap, 615, A47 \csname Subramanianlink\endcsname~\csname
  Subramaniannote\endcsname

\bibitem[{{Szenicer} {et~al.}(2019){Szenicer}, {Fouhey}, {Munoz-Jaramillo},
  {Wright}, {Thomas}, {Galvez}, {Jin}, \& {Cheung}}]{Szenicer}
{Szenicer}, A., {Fouhey}, D.~F., {Munoz-Jaramillo}, A., {et~al.} 2019, Science
  Advances, 5, eaaw6548 \csname Szenicerlink\endcsname~\csname
  Szenicernote\endcsname

\bibitem[{{Tajfirouze} {et~al.}(2016{\natexlab{a}}){Tajfirouze}, {Reale},
  {Peres}, \& {Testa}}]{Tajfirouze16}
{Tajfirouze}, E., {Reale}, F., {Peres}, G., \& {Testa}, P. 2016{\natexlab{a}},
  \apjl, 817, L11 \csname Tajfirouze16link\endcsname~\csname
  Tajfirouze16note\endcsname

\bibitem[{{Tajfirouze} {et~al.}(2016{\natexlab{b}}){Tajfirouze}, {Reale},
  {Petralia}, \& {Testa}}]{Tajfirouze}
{Tajfirouze}, E., {Reale}, F., {Petralia}, A., \& {Testa}, P.
  2016{\natexlab{b}}, \apj, 816, 12 \csname Tajfirouzelink\endcsname~\csname
  Tajfirouzenote\endcsname

\bibitem[{{Testa} {et~al.}(2013){Testa}, {De Pontieu}, {Mart{\'\i}nez-Sykora},
  {DeLuca}, {Hansteen}, {Cirtain}, {Winebarger}, {Golub}, {Kobayashi},
  {Korreck}, {Kuzin}, {Walsh}, {DeForest}, {Title}, \& {Weber}}]{Testa13}
{Testa}, P., {De Pontieu}, B., {Mart{\'\i}nez-Sykora}, J., {et~al.} 2013,
  \apjl, 770, L1 \csname Testa13link\endcsname~\csname Testa13note\endcsname

\bibitem[{{Thalmann} {et~al.}(2014){Thalmann}, {Tiwari}, \&
  {Wiegelmann}}]{Thalmann}
{Thalmann}, J.~K., {Tiwari}, S.~K., \& {Wiegelmann}, T. 2014, \apj, 780, 102
  \csname Thalmannlink\endcsname~\csname Thalmannnote\endcsname

\bibitem[{{Tian} {et~al.}(2014){Tian}, {DeLuca}, {Cranmer}, {De Pontieu},
  {Peter}, {Mart{\'\i}nez-Sykora}, {Golub}, {McKillop}, {Reeves}, {Miralles},
  {McCauley}, {Saar}, {Testa}, {Weber}, {Murphy}, {Lemen}, {Title}, {Boerner},
  {Hurlburt}, {Tarbell}, {Wuelser}, {Kleint}, {Kankelborg}, {Jaeggli},
  {Carlsson}, {Hansteen}, \& {McIntosh}}]{Tian14}
{Tian}, H., {DeLuca}, E.~E., {Cranmer}, S.~R., {et~al.} 2014, Science, 346,
  1255711 \csname Tian14link\endcsname~\csname Tian14note\endcsname

\bibitem[{{van Ballegooijen} {et~al.}(2017){van Ballegooijen}, {Asgari-Targhi},
  \& {Voss}}]{Van17}
{van Ballegooijen}, A.~A., {Asgari-Targhi}, M., \& {Voss}, A. 2017, \apj, 849,
  46 \csname Van17link\endcsname~\csname Van17note\endcsname

\bibitem[{{Viall} \& {Klimchuk}(2011)}]{Viall}
{Viall}, N.~M. \& {Klimchuk}, J.~A. 2011, \apj, 738, 24 \csname
  Vialllink\endcsname~\csname Viallnote\endcsname

\bibitem[{{Wang} {et~al.}(2019){Wang}, {Ugarte-Urra}, \& {Reep}}]{Wang}
{Wang}, Y.~M., {Ugarte-Urra}, I., \& {Reep}, J.~W. 2019, \apj, 885, 34 \csname
  Wanglink\endcsname~\csname Wangnote\endcsname

\bibitem[{{Wang} {et~al.}(2004){Wang}, {Bovik}, {Sheikh}, \&
  {Simoncelli}}]{Wang04}
{Wang}, Z., {Bovik}, A.~C., {Sheikh}, H.~R., \& {Simoncelli}, E.~P. 2004, IEEE
  Transactions on Image Processing, 13, 600 \csname
  Wang04link\endcsname~\csname Wang04note\endcsname

\bibitem[{{Warren} {et~al.}(2012){Warren}, {Winebarger}, \&
  {Brooks}}]{Warren12}
{Warren}, H.~P., {Winebarger}, A.~R., \& {Brooks}, D.~H. 2012, \apj, 759, 141
  \csname Warren12link\endcsname~\csname Warren12note\endcsname

\bibitem[{{Wheatland} {et~al.}(2000){Wheatland}, {Sturrock}, \&
  {Roumeliotis}}]{Wheatland}
{Wheatland}, M.~S., {Sturrock}, P.~A., \& {Roumeliotis}, G. 2000, \apj, 540,
  1150 \csname Wheatlandlink\endcsname~\csname Wheatlandnote\endcsname

\bibitem[{{Winebarger} {et~al.}(2013){Winebarger}, {Walsh}, {Moore}, {De
  Pontieu}, {Hansteen}, {Cirtain}, {Golub}, {Kobayashi}, {Korreck}, {DeForest},
  {Weber}, {Title}, \& {Kuzin}}]{Winebarger}
{Winebarger}, A.~R., {Walsh}, R.~W., {Moore}, R., {et~al.} 2013, \apj, 771, 21
  \csname Winebargerlink\endcsname~\csname Winebargernote\endcsname

\bibitem[{{Winebarger} \& {Warren}(2005)}]{Winebarger05}
{Winebarger}, A.~R. \& {Warren}, H.~P. 2005, \apj, 626, 543 \csname
  Winebarger05link\endcsname~\csname Winebarger05note\endcsname

\bibitem[{{Yang} {et~al.}(2018){Yang}, {Longcope}, {Ding}, \& {Guo}}]{Yang18}
{Yang}, K.~E., {Longcope}, D.~W., {Ding}, M.~D., \& {Guo}, Y. 2018, Nature
  Communications, 9, 692 \csname Yang18link\endcsname~\csname
  Yang18note\endcsname

\bibitem[{{Yu} {et~al.}(2018){Yu}, {Fan}, {Yang}, {Xu}, {Wang}, {Wang}, \&
  {Huang}}]{Yu}
{Yu}, J., {Fan}, Y., {Yang}, J., {et~al.} 2018, arXiv e-prints,
  arXiv:1808.08718 \csname Yulink\endcsname~\csname Yunote\endcsname

\bibitem[{{Yu} {et~al.}(2021){Yu}, {Xu}, \& {Yan}}]{Yu21}
{Yu}, X., {Xu}, L., \& {Yan}, Y. 2021, \solphys, 296, 56 \csname
  Yu21link\endcsname~\csname Yu21note\endcsname

\bibitem[{{Zhang} {et~al.}(2023){Zhang}, {Hou}, {Fang}, {Chen}, {Li}, {Yan},
  {Ding}, {Song}, {Xiang}, \& {Liu}}]{Zhang23}
{Zhang}, J., {Hou}, Y., {Fang}, Y., {et~al.} 2023, \apjl, 942, L2 \csname
  Zhang23link\endcsname~\csname Zhang23note\endcsname

\bibitem[{{Zhang} \& {Ni}(2019)}]{Zhang19}
{Zhang}, Q.~M. \& {Ni}, L. 2019, \apj, 870, 113 \csname
  Zhang19link\endcsname~\csname Zhang19note\endcsname

\end{thebibliography}

 \begin{appendix}

\section{ Estimation of width of a loop}\label{append:b}
\begin{figure}[hbtp]
  \centering
  \includegraphics[width=0.8\columnwidth]{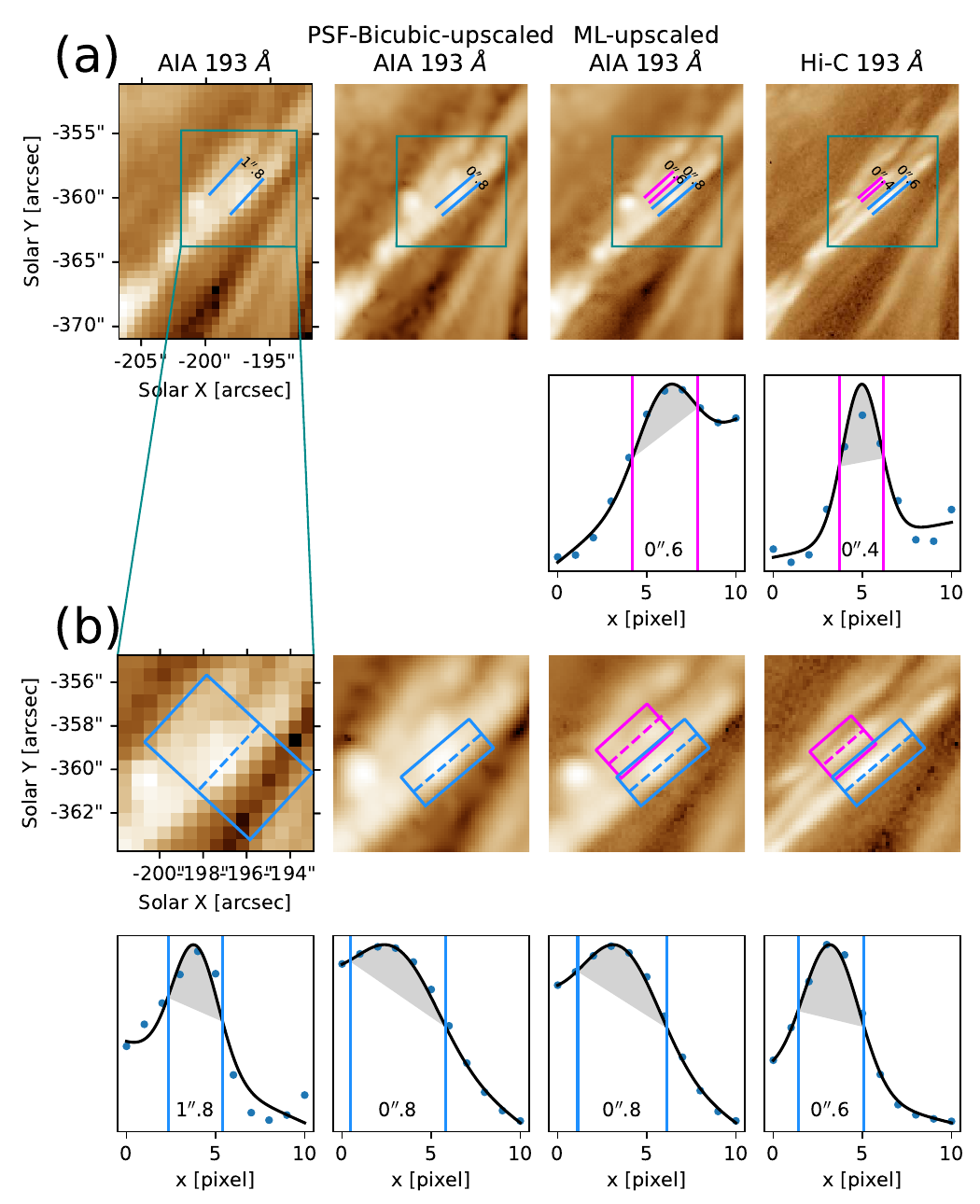}  
  \caption[]{\label{f9}   
  The estimated widths of the loops in  the various images taken from testing and validation sets. 
{\em Panel a:\/} Comparison of close-ups of the AIA, PSF-Bicubic-upscaled, ML-upscaled, and Hi-C images taken in the validation set.
{\em Panel b:\/} Comparison of more detailed close-ups of thees images.
The plots above and below each panel in {\em b} show  the distribution of the corresponding average intensity across the axis of the strip as indicated by the magenta and blue  dash line, respectively. 
In each plot, the curve refers to the fitted  gaussian function with the linear term; 
the distance between the vertical line corresponds to the  FWHM for the fitted function.
In {\em a}, each pair of  parallel lines  are deposited to cover the FWHM area for the corresponding fitting function as plotted in {\em b}.
  }
  \end{figure}

\begin{figure}[hbtp]
  \centering
  \includegraphics[width=0.8\columnwidth]{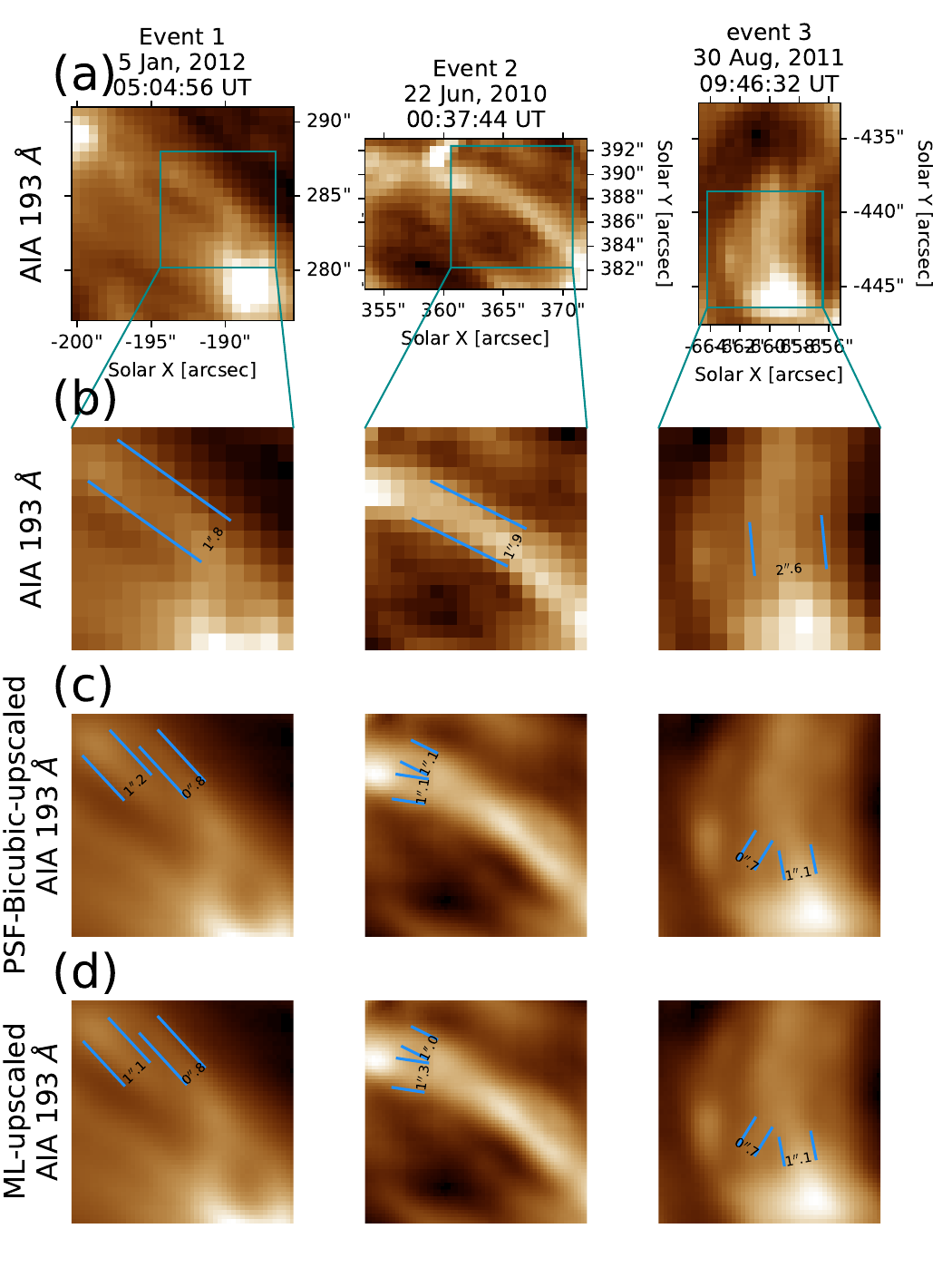}  
    \caption[]{\label{f10}   
  The estimated widths of the loops in  the various images taken from each event.
  {\em Panel a:\/} close-ups of the AIA 193 \AA\ images.
{\em Panel b:\/} more detailed close-ups of the AIA 193 \AA\ images.
{\em Panels c-d:\/}   the PSF-bicubic-upscaled and the  ML-upscaled AIA 193 \AA\ images with the same FOV as {\em b}, covering the braiding strands.
In each panel, the distance of the parallel lines colored the same is considered as the width of the corresponding loop or strand.
  }
  \end{figure}

We estimate the width of loops on various images as follows.
Firstly, we extract a straight stripe along a segment of the loop axis with a width of 11 pixels.
Secondly, we obtain the average intensity across the axis of the stripe, which is fitted to a Gaussian function with a linear term, expressed  as 
$f(x)=\frac{1}{\sqrt{2\pi}\sigma}e^{-(x-\mu)^{2}/2\sigma^{2}}+ax+b $
where $\sigma$, $\mu$, $a$ and, $b$ are fitting parameters. Thus, we define a full width at half maximum (FWHM) of $2.355\sigma$ as the width of a loop.
  In the validation set, as shown in Fig. \ref{f9}, a single AIA loop with  a width of 1{\arcsec}.8 appears to be two separated strands with the widths of 0{\arcsec}.6 and 0{\arcsec}.4 in the  simultaneous Hi-C image. 
 Similarly, the ML-upscaled AIA 193 \AA\  images  resolved the two strands, but slightly wider, with the width of 0{\arcsec}.6 and 0{\arcsec}.8, respectively.
 The strands   with the width  of 0{\arcsec}.8 was also recognizable on the PSF-Bicubic-upscaled AIA 193 \AA\ images.
 In each event reported here, the width of the resolved strands range from  0{\arcsec}.7 to 1{\arcsec}.3, which compose the  apparent  single AIA loops, with widths ranging from  1{\arcsec}.8 to 2{\arcsec}.6 (Fig. \ref{f10}).

    \end{appendix}

\end{document}